\documentclass[lettersize,journal]{IEEEtran}
\usepackage{amsmath,latexsym,amssymb,stmaryrd,pifont}
\usepackage{setspace,stackrel,tikz,graphicx}
\usepackage{exscale,relsize,subfig,textcomp,stackrel,setspace,float}
\usepackage{mdframed,pifont}
\usepackage{multicol,xfrac}
\usepackage{multirow, booktabs,placeins}
\usepackage{mathtools}
\usepackage{tabularx}
\usepackage[input-decimal-markers=.]{siunitx}
\usepackage{longtable}
\usepackage[T1]{fontenc}
\usepackage{graphicx}
\usepackage{caption}
\usepackage{fancyhdr}
\usepackage{lipsum}
\usepackage{multicol}
\usepackage{graphicx}
\usepackage[utf8]{inputenc}
\pdfoutput=1
\usepackage{amsthm}
\usepackage{array}
\usepackage{colortbl}
\usepackage{xcolor}
\usepackage{mathtools,amssymb,lipsum}
\usepackage{subfig}

\usepackage{cuted}
\usepackage{soul}
\usepackage[nolist]{acronym}

\begin{acronym}
    \acro{MIMO}{multiple-input, multiple-output}
    \acro{AWGN}{additive white Gaussian noise}
    \acro{HPBW}{half power beam width}
    \acro{TR}{technical report}
    \acro{KPI}{key performance indicator}
    \acro{RCS}{radar cross section}
    \acro{SCNR}{signal-to-clutter noise ratio}
    \acro{mmWave}{millimeter Wave}
    \acro{CI}{close-in}
    \acro{ISAC}{integrated sensing and communications}
    \acro{Tx}{transmitter}
    \acro{Rx}{receiver}
    \acro{PL}{path loss}
    \acro{PDF}{probability density function}
    \acro{CDF}{cumulative distribution function}
    \acro{GoF}{goodness-of-fit}
    \acro{3GPP}{3rd Generation Partnership Project}
    \acro{USRP}{universal software radio peripheral}
    \acro{InF}{indoor factory}
    \acro{InH}{indoor hotspot}
    \acro{UAV}{unmanned aerial vehicle}
    \acro{AMR}{autonomous mobile robot}
    \acro{RA}{robotic arm}
    \acro{LoS}{line-of-sight}
    \acro{H}{horizontal}
    \acro{KS}{Kolmogorov-Smirnov}
    \acro{NF}{near-field}
    \acro{MFE}{mean fitting error}
    \acro{MSE}{mean squared error}
    \acro{ZC}{Zadoff-Chu}
    \acro{IF}{intermediate frequency}
    \acro{QR}{quadruped robot}
    \acro{CIR}{channel impulse response}
    \acro{THz}{terahertz}
    \acro{FI}{floating intercept}
    \acro{RFIC}{radio-frequency integrated circuit}
    \acro{RF}{radio-frequency}
    \acro{NYU}{New York University}
    \acro{ADC}{analog-to-digital converter}
\end{acronym}

\begin{document}

\title{\vspace{0cm}\LARGE Indoor Statistical and Deterministic RCS Characterization for ISAC Channel Modeling\vspace{0em}}
\makeatletter
\patchcmd{\@maketitle}
  {\addvspace{0\baselineskip}\egroup}
  {\addvspace{0\baselineskip}\egroup}
  {}
  {}
\makeatother

\author{Ali Waqar Azim, Ahmad Bazzi, Roberto Bomfin, Nikolaos Giakoumidis, Theodore S. Rappaport, Marwa Chafii
\thanks{This work was supported in part by the Tamkeen through the Research Institute New York University Abu Dhabi (NYUAD) under Grant CG017; and in part by the NYUAD Center for Artificial Intelligence and Robotics, funded by Tamkeen through the Research Institute under Award CG010.\\
Ali Waqar Azim is with James Watt School of Engineering, University of Glasgow, G12 8QQ, UK. Ahmad Bazzi, Roberto Bomfin, and Marwa Chafii are with the Engineering Division, New York University Abu Dhabi (NYUAD), 129188, UAE
(email: aliwaqarazim@gmail.com,{ahmad.bazzi,roberto.bomfin, marwa.chafii}@nyu.edu). Ahmad Bazzi, Theodore S. Rappaport and Marwa Chafii are with NYU WIRELESS, NYU Tandon School of Engineering, Brooklyn, 11201, NY, USA (email: tsr@nyu.edu). Nikolaos Giakoumidis is with Center for AI and Robotics, NYUAD, 129188, UAE and with Intelligent Systems Lab, Cultural Technology and Communication, University of the Aegean, 811 00 Mitilini, Greece (email: giakoumidis@nyu.edu).}}
\maketitle
\begin{abstract}
In this study, we perform statistical radar cross section (RCS) analysis for various test targets in an indoor factory at \SI{25}{}-\SI{28}{\GHz}, with the goal of determining the best-fit parametric distributions that characterize the target scattering properties to be used in integrated sensing and communication channel modeling standardization. The analysis is conducted based on measurements in quasi-monostatic and bistatic configurations with bistatic angles of \(20^\circ\), \(40^\circ\), and \(60^\circ\). The test targets include unmanned aerial vehicles, an autonomous mobile robot, and a robotic arm. Goodness-of-fit tests validate that the RCS of these targets is best modeled by \textit{lognormal} and \textit{gamma} distributions with high statistical confidence. Additionally, we provide a framework for evaluating the \ac{NF}, specular-dominant effective bistatic RCS of a rectangular sheet under controlled bistatic geometries. Novel deterministic RCS models are evaluated, incorporating dependencies on the bistatic angle, transmitter-target separation (\SIrange{2}{10}{\meter}). The results demonstrate that some proposed deterministic RCS models accurately fit the measured data, highlighting their applicability in deterministic RCS characterization in NF bistatic configurations.
\end{abstract}
\IEEEpeerreviewmaketitle
\vspace{-4mm}
\section{Introduction}
\Ac{ISAC} paradigm integrates wireless communications and radar sensing \cite{11519592,11173662} to enable joint use of spectrum and hardware, typically in monostatic or bistatic configurations \cite{9449071}. In the context of \ac{ISAC} channel modeling \cite{11574587}, the \ac{RCS} is one of the fundamental target descriptor that quantifies its scattering strength and provides a physically meaningful basis for distinguishing targets of different size, material, and structural complexity. The \ac{RCS}, in fact, defines an effective reflective area by relating the incident field to the backscattered energy received by the sensing system, directly impacting the targets' detectability in radar-based sensing within \ac{ISAC} \cite{wei2024integrated, chowdary2024hybrid,108383}.

The \ac{RCS} of a sensing target can be characterized using deterministic (e.g., site-specific \cite{108383}) or statistical modeling approaches \cite{wei2024integrated}. Deterministic \ac{RCS} modeling relies on numerical solvers to compute \ac{RCS} from the target’s geometry and material properties. On the other hand, statistical \ac{RCS} modeling represents \ac{RCS} as a random variable governed by a parametric \ac{PDF} \cite{borden1983statistical}. The formulation is particularly appropriate when the power of the reflected echo from the target fluctuates due to uncontrolled variations in its aspect, surface scattering mechanisms, etc. In this case, measured \ac{RCS} realizations are fitted to candidate parametric distributions, and the most suitable model is selected using \ac{GoF} criteria, yielding compact distribution parameters that are directly usable in system-level simulators. However, in case we know the geometry and \ac{NF} regimes, \ac{RCS} is no longer well represented by a distance-invariant random variable. Instead, \ac{RCS} exhibits systematic dependence on range and bistatic angle (in bistatic configuration). In such cases, deterministic (configuration-aware) \ac{RCS} parameterizations are required for digital-twin/map-based evaluation and for diagnosing angular/range signatures.

Given the emergence of high-bandwidth, narrow beamforming technologies, the \ac{3GPP} has adopted statistical \ac{RCS} modeling for \ac{ISAC} channel modeling
\cite{3gpp_r1_2408100}. Accordingly, industrial contributions have discussed several candidate distribution families, most
notably lognormal, Weibull, and gamma, as viable models for target \ac{RCS} statistics. More recently, \ac{TR}38.901 \cite{3gpp38901} consolidates this effort by adopting a lognormal-based \ac{RCS} representation for standardized target classes by proposing a three parameter based \ac{RCS} model following a two-stage methodology: (i) estimating distribution parameters through statistical fitting of measured \ac{RCS} data, and (ii) aggregating the resulting parameters into a compact, frequency-agnostic representation suitable for standardized use.

Within \ac{TR}38.901 \cite{3gpp38901}, the sensing-target contribution is modeled as an additional \textit{target channel} component appended to the conventional \textit{background channel}. The target impact is parameterized primarily through \ac{RCS} as: (i) an amplitude-scaling descriptor for target channel components and (ii) \ac{PL} formulation used for the target channel. In this context, the best-fit \ac{RCS} distributions and estimated parameters reported in this work provide measurement-grounded inputs for incorporating the target \ac{RCS} block for the considered target classes and sensing geometries. Moreover, the extracted parameters listed herein for the lognormal distribution can be used to derive the three parameters required for the \ac{TR}38.901 \ac{RCS} model.
\vspace{-3mm}
\subsection{Literature Review}
Recent advances in radar-based sensing have led to a significant body of literature addressing \ac{RCS} modeling, measurement, and statistical characterization at different operating frequencies and target classes. \cite{ezuma2021radar} present a classification system based on measured \ac{RCS} for \acp{UAV} at \SI{15} and \SI{25}{\GHz} where the \ac{RCS} is shown to depend on the shape, material, azimuth angle, frequency, and polarization of the \ac{UAV}. Their analysis demonstrates that lognormal, gamma, and generalized extreme value distributions model \ac{RCS} most accurately, outperforming Gaussian assumptions. In a related effort, \cite{rosamilia2022radar} conducted \ac{RCS} measurements for \acp{UAV} in the \SIrange{8.2}{18}{\GHz}.  \cite{ezuma2022comparative} benchmarked numerous machine learning and deep learning algorithms for \ac{UAV} recognition using \ac{RCS} data measured at \SI{15} and \SI{25}{\GHz}. A broader overview is offered in the review work by \cite{patel2018review}, which surveys existing radar classification and \ac{RCS} modeling techniques for small-sized \acp{UAV}. Additional measurement efforts, such as those by \cite{semkin2021drone}, collected \SI{28}{\GHz} \ac{RCS} data of carbon fiber \acp{UAV} and applied the measurements to assess detection probabilities in simulated urban environments. Similarly, \cite{semkin2020analyzing} provided open-access \ac{RCS} data of different \ac{UAV} models between \SIrange{26}{40}{\GHz}, confirming that material composition and target size significantly influence radar visibility. In \cite{azim2025rcs}, the \ac{3GPP}-proposed three-parameter lognormal \ac{RCS} model was employed to estimate the corresponding model parameters for various targets operating in the \SI{25}{}-\SI{28}{\GHz} frequency range. Beyond measurement-centric studies, recent \ac{ISAC} optimization works have started to explicitly account for \ac{RCS} uncertainty, e.g., by modeling dynamic/look-angle-dependent \ac{RCS} as a random variable within chance-constrained quality-of-service formulations and outage-based sensing reliability metrics in movable-antenna assisted \ac{ISAC} designs \cite{khalili2024advanced,khalili2025pinching}. Existing literature reveals that prior research has not systematically investigated statistical \ac{RCS} modeling for other targets, such as \ac{RA} and \ac{AMR} in \ac{InF} environments through rigorous measurements and testing of multiple candidate parametric distributions.
\vspace{-2mm}
\subsection{Contributions}
Our study characterizes the \ac{RCS} of different targets using  statistical and deterministic modeling approaches in both \ac{InF} and \ac{InH} environments.
For the statistical \ac{RCS} characterization, we conduct a measurements in an \ac{InF} environment over the \SI{25}{}-\SI{28}{\GHz} and quantify \ac{RCS} variations under quasi-monostatic and bistatic sensing geometries. The targets are common \ac{InF} objects, including two \acp{UAV} of different size and material composition; robotic platforms, i.e., an \ac{AMR} and a \ac{RA}. We fit four candidate distribution families, normal, lognormal, gamma, and Weibull, consistent with the distributions considered in industrial contributions to \ac{3GPP} \ac{ISAC} channel modeling standardization \cite{3gpp_r1_2408100} to the measured \ac{RCS} data. The best fit distribution is selected using \ac{GoF} criteria based on the \ac{KS} statistic and \ac{MSE}. The  distribution parameters are reported for generic target classes (type/size categories) rather than exact geometries, emphasizing reproducible scattering behavior over precise physical specifications to be used in \ac{TR}38.901 style \ac{ISAC} simulators for target dependent amplitude statistics.

In addition, here we present a deterministic framework for \ac{NF} bistatic \ac{RCS} characterization of a rectangular sheet in \ac{InH} over \SI{25}{}-\SI{28}{\GHz}. The approach leverages \ac{PL} measurements across different  bistatic angles, $\theta_b$ to derive an angle and range dependent \ac{RCS} model and to estimate propagation parameters including the \ac{PL} exponent, intercept, and shadowing factor. The formulation follows a modified \ac{FI}  and \ac{CI} (with \SI{1}{\meter} reference distance) models that incorporate two-way \ac{PL} together with an explicit \ac{RCS} term. The optimal model is selected using \ac{MFE}.

To summarize, this study makes the following contributions:
\begin{enumerate}
\item For statistical \ac{RCS} characterization, we determine parametric distributions that best fit the empirical \ac{RCS} for different \ac{InF} targets through systematic measurements over \SI{25}{}-\SI{28}{\GHz} band under quasi-monostatic and bistatic configurations with \(\theta_b \in \left\{20^\circ,40^\circ,60^\circ\right\}\). To the best of our knowledge, there is no study available in the literature that has done the same.
\item We develop and validate a deterministic near field and specular-dominant effective \ac{RCS} model for a rectangular plate under a controlled \ac{Tx}-target-\ac{Rx} geometry. The model captures the dependence on \(\theta_b\) through empirically derived polynomial coefficients in the \SI{25}{}-\SI{28}{\GHz} band and provides a configuration-aware \ac{RCS} term for site-specific or \ac{NF} channel modeling when the received target contribution is dominated by the specular bistatic return.
\end{enumerate}
\subsection{Paper Organization}
The remainder of the paper is organized as follows. Section \ref{sec2} describes the measurement campaign, including the setup, equipment, targets for statistical and deterministic \ac{RCS} characterization. In addition, it also provides the measurement scope and domain of validity of the presented results. Section \ref{sec3} presents the statistical \ac{RCS} analysis, covering the processing methodology (Section~\ref{sec3a}), the candidate distributions and \ac{GoF} tests (Section~\ref{sec3b}), and the resulting distribution fits (Section~\ref{sec3c}). Section \ref{sec4} develops the \ac{NF} deterministic bistatic \ac{RCS} model based on \ac{PL} measurements in the \ac{InH} environment. Section~\ref{sec5} concludes the paper.
\section{Measurement Campaign: Setup, Equipment, and Targets}\label{sec2}
\subsection{Measurement Setup and Environment}\label{sec2a}
\begin{figure}[t]
    \centering
    \includegraphics[width=1.02\linewidth]{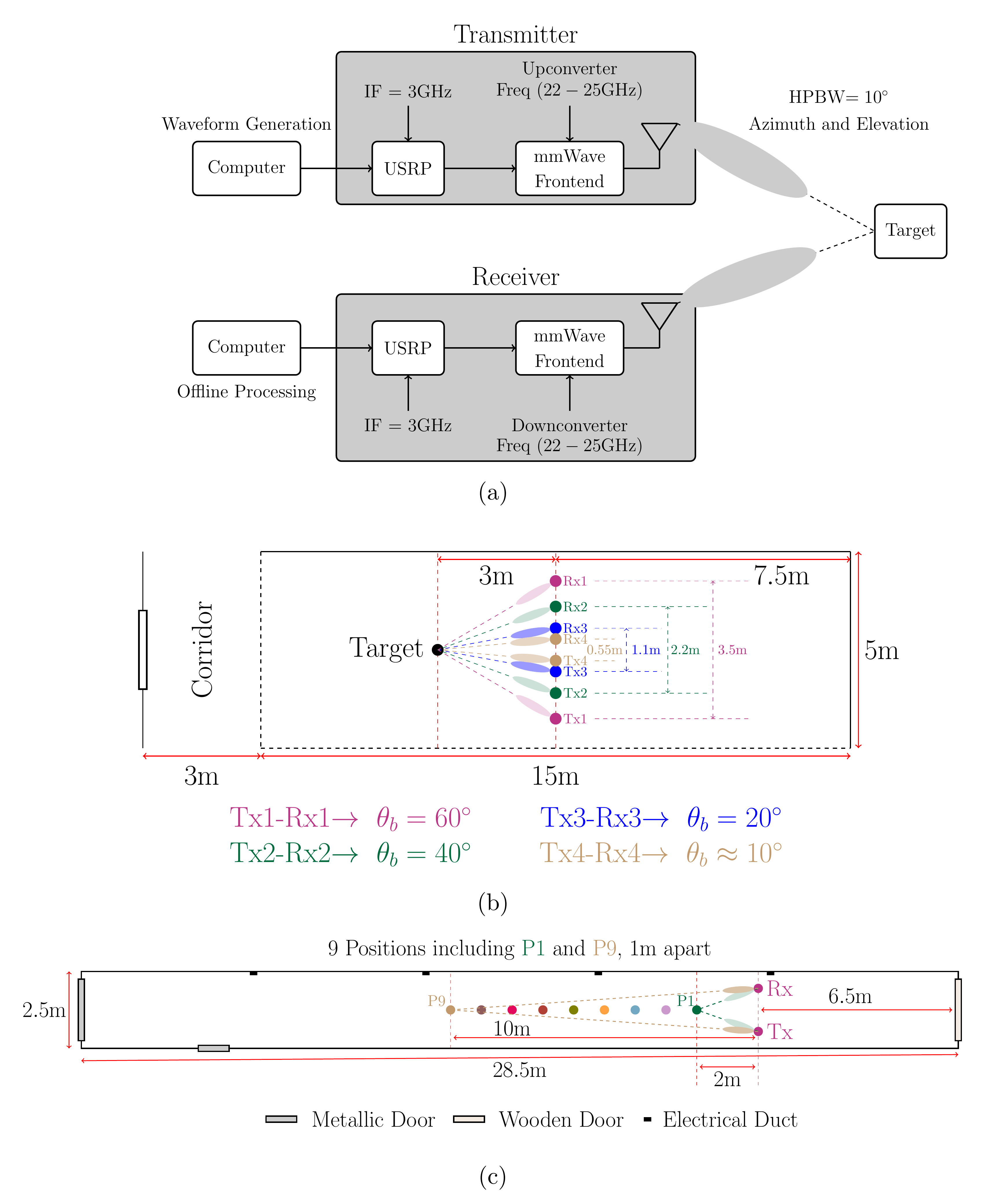} 
    \caption{(a) Testbed for the RCS measurement, (b) the   layout of the measurement environment for statistical RCS analysis, (c)  the   layout of the measurement environment for deterministic RCS analysis.  }
    \label{fig1} 
\end{figure}
\textcolor{black}{The block diagram of our measurement testbed is illustrated in Fig. {\ref{fig1}(a)}, which shows the} bistatic \textcolor{black}{configuration} used for \ac{RCS} measurements. \textcolor{black}{The testbed comprises signal generation, upconversion, transmission, and corresponding symmetric reception processes.} The system uses a \ac{USRP} to digitize \textcolor{black}{the waveform generated offline using MATLAB}, \ac{mmWave} front-end \acp{RFIC} for frequency upconversion and downconversion. Fig. {\ref{fig1}(b)} provides the top-view of the statistical \textcolor{black}{\ac{RCS}} measurement setup showing four distinct \ac{Tx}-\ac{Rx} pairs positioned at different \(\theta_b\) with the boresight of the antenna tilted towards the target. \ac{Tx}/\ac{Rx} alignment was done using a crossed-laser system, where two laser beams intersected at the \textit{observation point} \textcolor{black}{located \SI{1}{\meter} above ground level and \SI{3}{\meter} away} from the center of the line segment connecting the \ac{Tx} and the \ac{Rx}. The geometric center of each target was manually positioned at this laser-defined reference point, verified through laser grid projection. The \ac{Tx} and the \ac{Rx} are placed in the middle of the \textcolor{black}{measurement space} which includes dense clutter characteristic of typical \ac{InF} scenarios, ensuring realistic propagation conditions. The laboratory environment where the measurements were conducted was designed to emulate propagation characteristics \textcolor{black}{of \ac{InF}, e.g., high ceiling height, etc.,} while maintaining controlled conditions. \textcolor{black}{The measurement area is housed in the KINESIS Lab within NYU Abu Dhabi's Core Technology Platforms, situated on Saadiyat Island in the UAE. The facility measures \SI{5}{\meter}\(\times\)\SI{15}{\meter}\(\times\)\SI{8.5}{\meter} and is located \textcolor{black}{on the second basement level of the} university's Arts Center.}

\textcolor{black}{Fig. \ref{fig1}(c) shows the top-view layout of the \ac{InH} corridor, with dimensions \SI{2.5}{\meter}\(\times\) \SI{28.5}{\meter}\(\times\) \SI{2.5}{\meter}, which is employed for deterministic \ac{RCS} measurements.} The target traverses nine predefined positions (P1-P9) to systematically vary \textcolor{black}{\(\theta_b\)}. For precise alignment, \textcolor{black}{the sounding} system employs crossed laser markers, with their intersection point calibrated to the target's geometric center. \textcolor{black}{The \ac{Tx} and \ac{Rx} are positioned on a line located \SI{6.5}{\meter} away from one end of the layout.}

\textcolor{black}{All measurements were conducted under \ac{LoS} propagation conditions, employing \ac{H} polarization at both the \ac{Tx} and \ac{Rx}. Both measurement setups were kept unchanged for the entire campaign, ensuring that the collected data is not affected by artifacts.} The total amount of data collected throughout our campaign is \(3.544\) gigabytes.
\subsection{Measurement Equipment}\label{sec2b}
\textcolor{black}{The \ac{Tx} and \ac{Rx} are equipped with Sivers \(\mathrm{EVK02004}\) \acp{RFIC}, each featuring a \(4\times 4\) antenna array that provides a \(10^\circ\) \ac{HPBW} in both the azimuth and elevation planes.} Each \ac{RFIC} is interfaced with \(\mathrm{B205mini}\) \ac{USRP} with \SI{20}{\MHz} bandwidth. Measurements were conducted at \SI{25}{}-\SI{28}{\GHz}, each with a \SI{20}{\MHz} instantaneous bandwidth centered on these frequencies. The \acp{USRP} operate at an intermediate frequency of \SI{3}{\GHz}. At the \ac{Tx}, \ac{ZC} sequence of length \(128\) was generated at baseband and transmitted directly through the \ac{USRP}, occupying its full \SI{20}{\MHz} bandwidth confirmed by the subsequent spectrum measurements. The analog-to-digital converter of the \ac{USRP} has a resolution of $12$ bits. Frequency tuning of the \ac{RFIC} is done by adjusting the upconverter and downconverter frequencies in \SI{1}{\GHz} increments \textcolor{black}{followed by a calibration step} to ensure optimal performance at the new operating frequency.
\subsection{\textcolor{black}{Measurement Scope and Domain of Validity}}\label{sec2c}
\textcolor{black}{The reported statistical and deterministic \ac{RCS} characterization is performed under a controlled boresight-illumination validity envelope, i.e., the target response is estimated while it remains within the \ac{Tx}/\ac{Rx} main beams and in \ac{LoS} propagation, which is particularly important in our testbed, as the \ac{Tx}/\ac{Rx} employ highly directive arrays with \ac{HPBW} of $\approx 10^\circ$ in both azimuth and elevation. All measurements follow this same envelope: (i) fixed and calibrated \ac{Tx}/\ac{Rx} placement, (ii) controlled target placement around the observation point (for statistical \ac{RCS} characterization) and pre-defined points (for deterministic \ac{RCS} characterization), and (iii) motion protocols designed to keep the target within the main lobes of both \ac{Tx} and \ac{Rx} (in case of statistical \ac{RCS} characterization).} 

\textcolor{black}{For statistical \ac{RCS} characterization, a small set of \(\theta_b\) with the target located at the observation point is considered. Therefore, the fitted distribution parameters should be interpreted as valid for the considered \textit{boresight-aligned geometries} rather than as scenario-agnostic constants. Any significant off-axis translation or combined translation/rotation bias the inferred \ac{RCS} due to beam misalignment and is therefore excluded from the present campaign. Furthermore, the reported statistical \ac{RCS} models are \textit{snapshot amplitude statistics} (\ac{PDF}/\ac{CDF}) that are \textit{aspect/pose/time-marginal}. Note that this work does not explicitly parameterize any Doppler signatures nor parameterize \ac{RCS} versus pitch/roll, but rather reports marginal \ac{RCS} \textit{amplitude} distributions obtained from target-only power extraction. Thus, any micro-Doppler signatures are inherently absorbed into the empirical distribution of the extracted target power and the resulting \ac{RCS}.}

The deterministic \ac{RCS} framework is \textit{validated for the measured rectangular sheet only}. Since the campaign includes a single sheet size, the polynomial coefficient-to-size mapping is under-identified. Hence, we do not claim a universal scaling \ac{RCS} law across sheet sizes from the present dataset. The fitted coefficients should therefore be interpreted as effective geometry-dependent parameters for the tested sheet. \textcolor{black}{Note that the deterministic rectangular-sheet model is not intended to represent a complete generalized bistatic \ac{RCS} function over independently varying incident and scattered directions. In the deterministic campaign, the sheet is used as a canonical planar reflector and is oriented such that the receiver observes the dominant specular bistatic return under the controlled Tx-target-Rx geometry. Therefore, the proposed model should be interpreted as a \ac{NF}, specular-dominant, configuration-aware effective \ac{RCS} parameterization, rather than as an off-specular or arbitrary multi-static scattering model. Receiver locations away from the specular branch would require additional angular scattering measurements or electromagnetic scattering models, which are outside the scope of this work. However, the fitting model in (11) generalizes the near-field specular RCS estimation across a continuous range of bistatic angles ($\theta_b$) and link distances within the measured domain.}
\subsection{Targets}\label{sec2d}
\subsubsection{Statistical \ac{RCS} Characterization Targets}
These targets are commonly found in \ac{InF} environments, which include: (i) two types of \acp{UAV}, (ii) a \ac{RA}, and (iii) an \ac{AMR}, referred to as the \ac{QR}. \textcolor{black}{Aligning with the \ac{3GPP} approach, the chosen targets reflect a range of categories rather than a comprehensive sampling of all possible shapes, sizes, and materials. The specifications of the targets are provided in Table \ref{equipment_table}. For all targets, aspect diversity is obtained by executing a prescribed motion/pose protocol while recording discrete-time channel snapshots. The target \emph{aspect/pose state} denoted by $\boldsymbol{\xi}(t)$ parametrizes the instantaneous scattering configuration (e.g., azimuth/elevation angles, altitude, joint-space pose, or discrete aspect labels). Accordingly, as aforementioned, the statistical \ac{RCS} results provided hereby should be interpreted as aspect-/pose-marginal distributions obtained by sweeping $\boldsymbol{\xi}(t)$ during the measurement protocol rather than as aspect-conditioned \ac{RCS}.}

\begin{figure}[t]
    \centering
    \subfloat[Mavic \(2\) Pro]{
        \includegraphics[width=0.45\columnwidth]{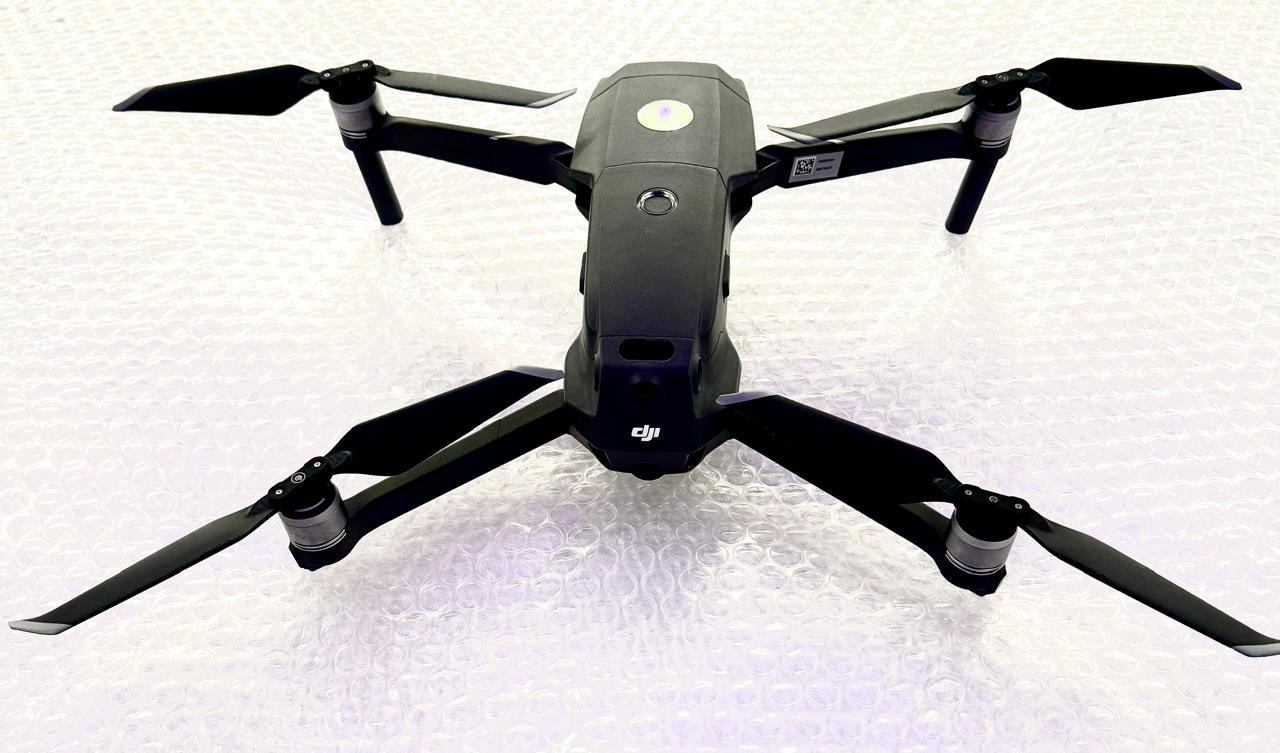}
        \label{drone01}
    }
    \hfill
    \subfloat[Matrice \(300\) RTK]{
        \includegraphics[width=0.43\columnwidth, height=2.4cm, keepaspectratio]{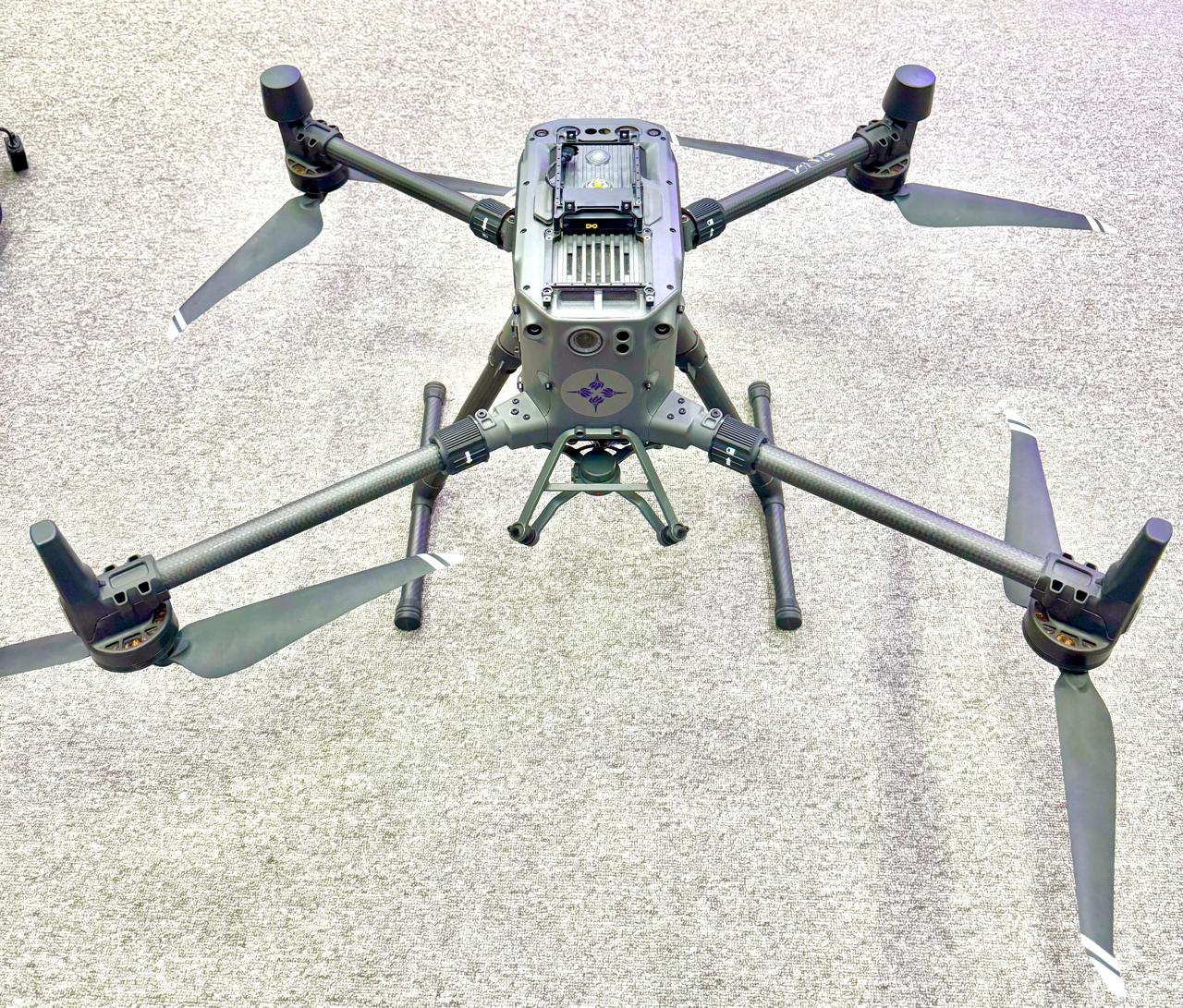}
        \label{drone02}
    }
    \caption{UAV test targets used in the study.}
    \label{drones}
\end{figure}

\textcolor{black}{The two \ac{UAV} platforms considered are the Mavic \(2\) Pro and the Matrice \(300\) RTK (cf. Fig. \ref{drones}). The folded/unfolded dimensions of the two \acp{UAV} are summarized in Table \ref{equipment_table}. For \acp{UAV}, $\boldsymbol{\xi}(t)$ is dominated by the yaw angle under nominal hover. Measurements were conducted with the platforms in stable hover while executing controlled yaw rotations about the observation point. Positional stability was maintained using crossed laser markers whose intersection was fixed at \SI{1}{\meter} above ground level, and drift was minimized to preserve boresight alignment. The vertical clearance from ground to the bottom of the platform differs due to geometry, i.e., approximately \(0.9\) m for the Mavic \(2\) Pro and \(0.6\) m for the Matrice \(300\) RTK. In addition, the battery placement also differs between platforms (side-mounted for the Matrice \(300\) RTK versus top-mounted for the Mavic \(2\) Pro), which may alter dominant scattering centers.}

\begin{figure}[t]
    \centering
    \subfloat[Robotic Arm]{
        \includegraphics[width=0.6\columnwidth, height=4.6cm, keepaspectratio]{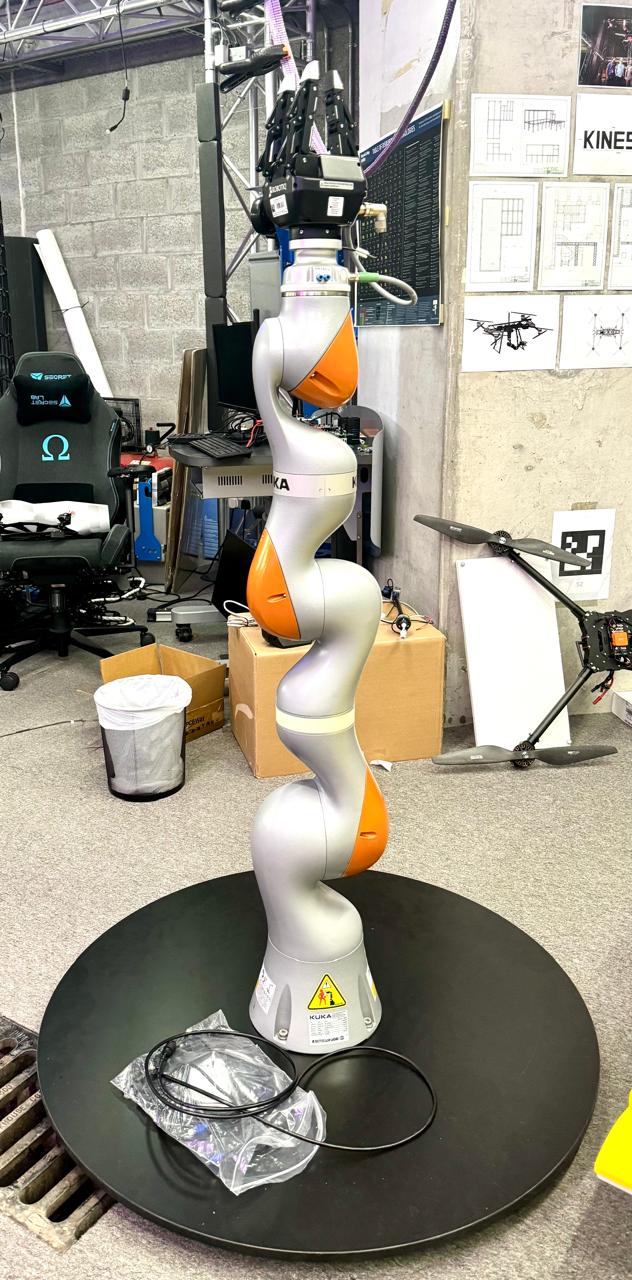}
        \label{robotic_arm}
    }
    \hfill
    \subfloat[Quadruped Robot]{
        \includegraphics[width=0.43\columnwidth]{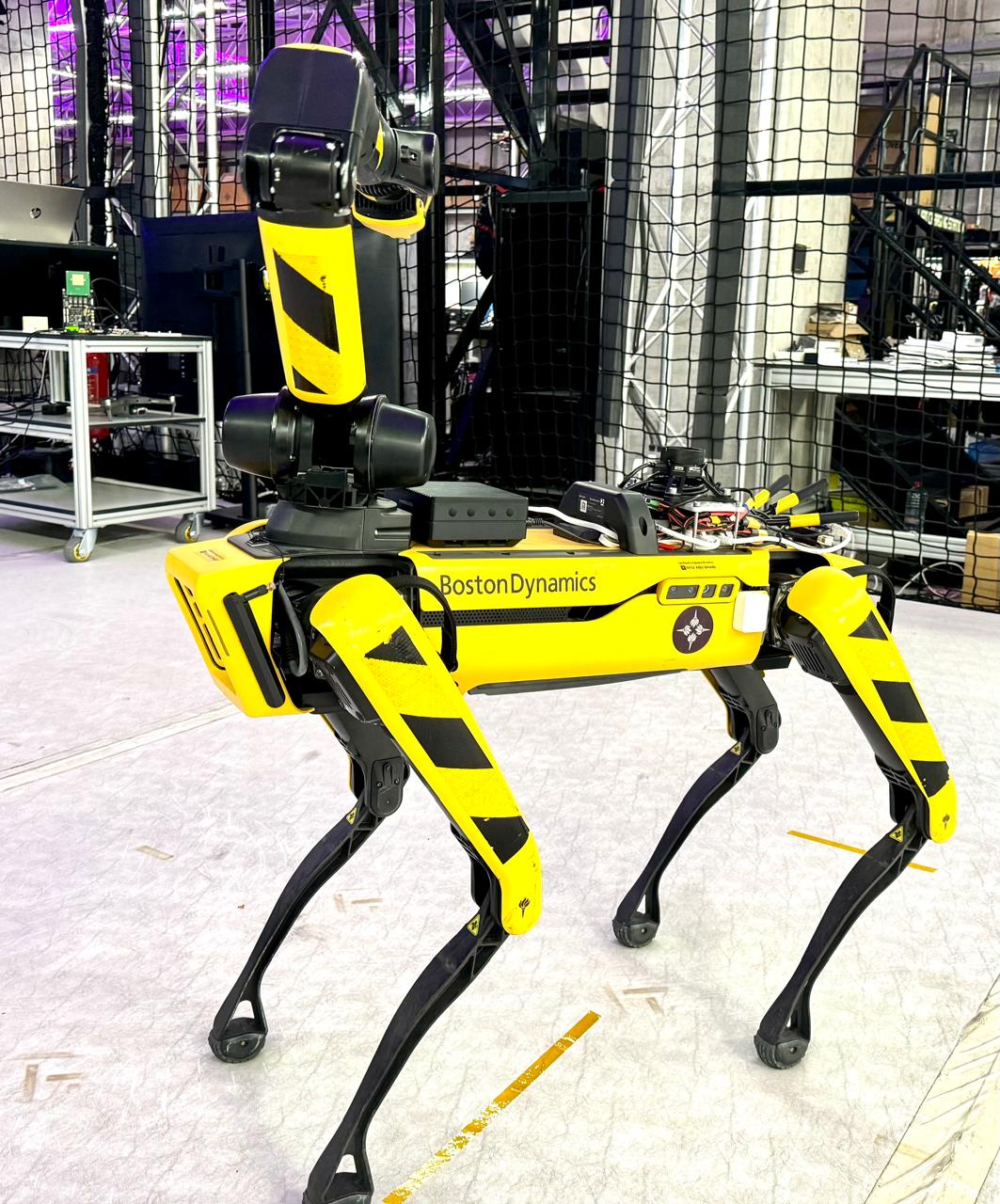}
        \label{robot}
    }
    \caption{InF equipment used in the study: (a) RA, (b) QR.}
    \label{targets}
\end{figure}

\textcolor{black}{The second target is the \ac{RA} (cf. Fig. \ref{robotic_arm}), whose reach and height are summarized in Table \ref{equipment_table}. For the \ac{RA}, $\boldsymbol{\xi}(t)$ corresponds to its articulated pose (seven degree-of-freedom joint configuration). Measurements were conducted with the \ac{RA} positioned at the observation point and continuously articulated through representative workspace motions, e.g., pick-and-place type trajectories, thereby generating \ac{RCS} samples over a broad set of aspect micro-states without imposing a fixed azimuth/elevation step size. Note that} the \ac{RA}'s maximum achievable height is \(1640\) mm. However, it varies during specific motions, such as those that simulate pick-and-place tasks, thus decreasing the reflective surface area. 

\textcolor{black}{The third target is the \ac{QR}  as shown in Fig.~\ref{robot}. The dimensions are specified in Table \ref{equipment_table}. In case of \ac{QR}, $\boldsymbol{\xi}(t)$ comprises a combination of discrete azimuth aspects and controlled motion primitives relative to the observation point. The \ac{RCS} data is collected during lateral motion (side-profile scattering) and longitudinal motion toward/away from the observation point (front/back scattering). In addition, explicit coarse azimuthal profiling was performed by sequentially measuring the front (\(0^\circ\)), lateral (\(90^\circ\)), and back (\(180^\circ\)) aspects with bilateral symmetry used for the two lateral sides. To probe elevation-dependent scattering under fixed boresight illumination, the \ac{QR} height was varied using a \SI{20}{\centi\meter} non-conductive platform.}
\begin{table*}[ht]
\small 
\centering
\caption{Specifications of Equipment Used in Measurements}
\begin{tabular}{|p{3.7cm}|p{5cm}|p{1.7cm}|p{5cm}|}
\hline
\textbf{Equipment} & \textbf{Dimensions} & \textbf{Mass} & \textbf{Materials} \\ \hline
DJI Mavic \(2\) Pro & Folded: \(214\times 91 \times 84\)mm & \(907\)g & Magnesium alloy, reinforced plastic, \\
&Unfolded: \(322 \times 242 \times 84\) mm&&carbon fiber, glass, silicon components\\\hline
DJI Matrice \(300\) RTK & Folded: \(430\times 420 \times 430\) mm & \(3.6\)-\(6.3\)kg & Carbon fiber-reinforced plastic, \\ 
& Unfolded: \(810 \times 670 \times 430\) mm&&aluminum, plastic \\\hline
Robotic Arm & Reach: \(820\) mm & \(\approx 30\)kg & Aluminum, titanium, steel, plastic, \\ 
(KUKA LBR iiwa 14 R\(820\)) &Height: \(1640\) mm (with arm)&& polymer composites\\\hline
Quadruped Robot & Length: \(1100\) mm; Width: \(500\) mm; & \(\approx 32.7\)kg & Aluminum, titanium, carbon fiber  \\ 
(Boston Dynamics Spot)& Height: \(610\) mm (walking without arm) &&composites, polymer\\\hline
\end{tabular}
\label{equipment_table}
\end{table*}
%
\subsubsection{Deterministic RCS Characterization Target} 
We employ a \SI{1.84}{\meter}\(\times\)\SI{1.2}{\meter} laminated wood rectangular sheet as a canonical planar reflector (target) to extract the \ac{NF} bistatic \ac{RCS} with known Tx-target-Rx positions (known geometry) that captures the dominant specular facet return versus range and \(\theta_b\). The selected laminated wooden sheet provides a mechanically rigid and repeatable planar reflector, enabling isolation of the systematic \textit{angle-/range-dependent} scattering behavior under \ac{NF} bistatic illumination. Many industrial objects, e.g., robot chassis panels, protective covers, machine enclosures, doors, ducts, and structural facades contain locally planar facets that exhibit specular dominant scattering. In such cases, a complex target can be approximated by a set of dominant planar facets.
\section{Statistical RCS Characterization of Indoor Factory Targets}\label{sec3}
In this section, we provide the statistical RCS evaluation methodology and the fitting of the chosen parametric distributions \textcolor{black}{to} the measured RCS data. 
\subsection{Experimental Setup and RCS Evaluation Methodology}\label{sec3a}
\begin{figure}[t]
    \centering
    \includegraphics[width=0.4\textwidth]{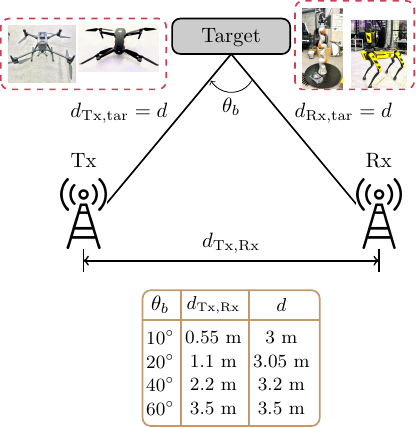} 
    \caption{An illustration of the measurement setup configuration.}
    \label{measurement_setup} 
\end{figure}

\begin{figure}[t]
    \centering
    \includegraphics[width=0.45\textwidth]{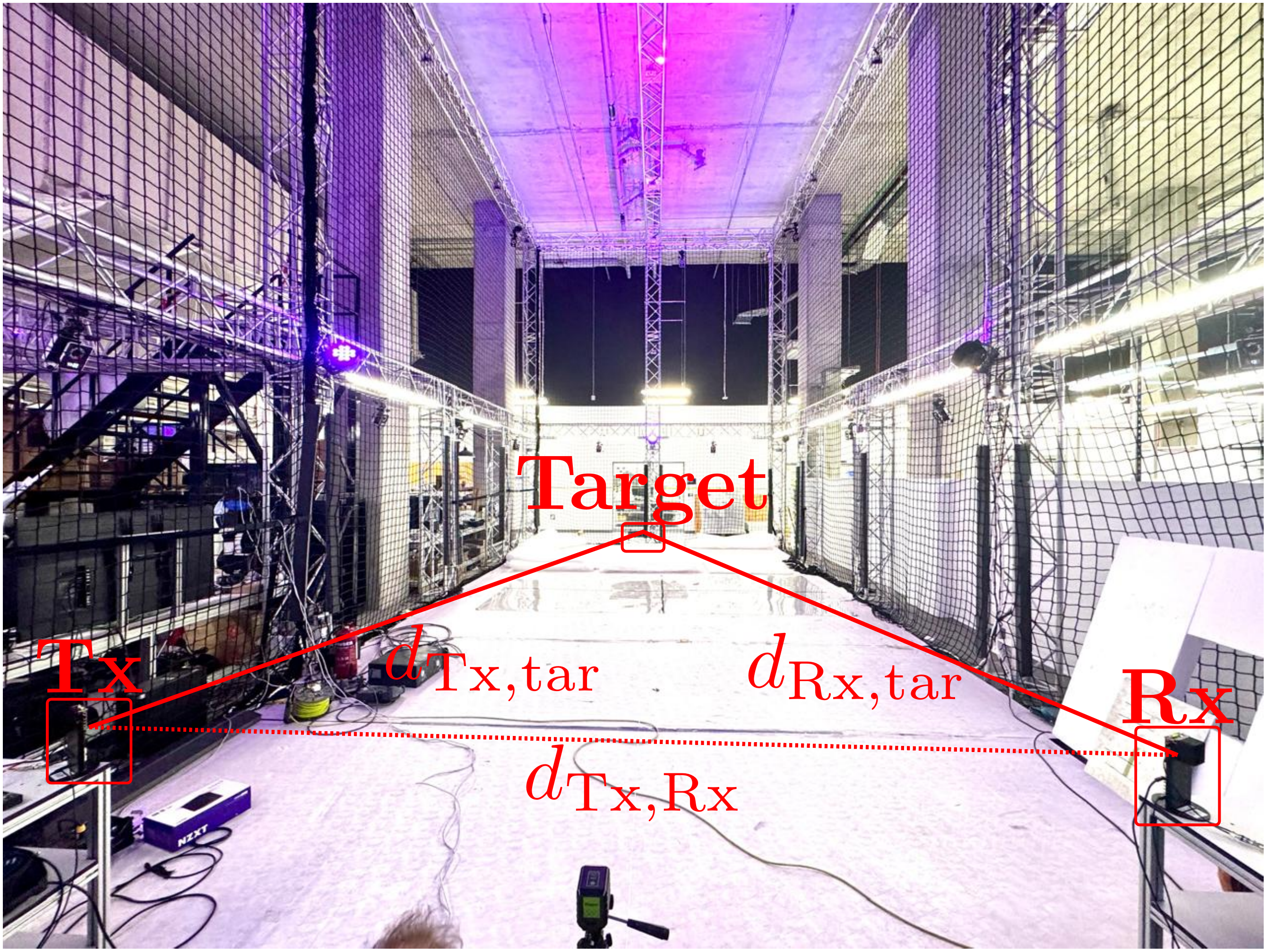} 
    \caption{The {\ac{InF}} environment and the measurement setup.}
    \label{actual_setup} 
\end{figure}

\textcolor{black}{
Geometry of the experimental setup is shown in Fig. \ref{measurement_setup} along with distances that result in quasi-monostatic and the three bistatic configurations with $\theta_b \in \{20^\circ,40^\circ,60^\circ\}$. Since monostatic configuration with $\theta_b=0^\circ$ poses practical limitation due to \ac{Tx}/\ac{Rx} self-interference,  therefore, we realize a quasi-monostatic setup by separating \ac{Tx} and \ac{Rx} by \SI{55}{\cm}, yielding $\theta_b \approx 10^\circ$. The \ac{Tx}-\ac{Rx}, \ac{Tx}-target, and \ac{Rx}-target distances are denoted by $d_{\mathrm{Tx,Rx}}$, $d_{\mathrm{Tx,tar}}$, and $d_{\mathrm{Rx,tar}}$, respectively, where $d_{\mathrm{Tx,tar}}=d_{\mathrm{Rx,tar}}=d$ is maintained during the measurement campaign. The corresponding \ac{InF} environment and the deployed geometry is shown in Fig. \ref{actual_setup}.}

\textcolor{black}{At the \ac{Rx}, the \ac{ZC}-based sounding signal is processed to estimate the \ac{CIR} for successive snapshots, which encapsulates information regarding the propagation environment, presence or absence of a target, and scattering characteristics within the channel. To enable robust target discrimination, the setup is operated at sufficiently high \ac{SCNR} such that relevant scatterers remain above the noise floor. For each $(f,\theta_b)$ configuration, we first acquire \textit{background} \acp{CIR} in the absence of the target, denoted by $h_{\mathrm{back}}(n)$, which encapsulate static clutter and the intrinsic response of the measurement domain. During the target measurements, i.e., measurement with target at the observation point, we estimate the snapshot-wise \ac{CIR}, $h_k(n)$ and compute the total received energy $P_{\mathrm{tot},k}=\sum_n |h_k(n)|^2$. Subsequently, using the pre-calculated background power
$P_{\mathrm{back}}=\sum_n |h_{\mathrm{back}}(n)|^2$ and the noise power $P_{\mathrm{noise}}$, we isolate the target-induced power per snapshot as
$P_{\mathrm{tar},k}=P_{\mathrm{tot},k}-P_{\mathrm{back}}-P_{\mathrm{noise}}$. The snapshot-wise formulation is used for all targets for \ac{RCS} extraction.}

\textcolor{black}{Considering the geometry, i.e., \(d = d_{\mathrm{Tx,tar}} = d_{\mathrm{Rx,tar}}\), the target-induced power relates to the instantaneous \ac{RCS} by the calibrated radar equation:
\begin{equation}\label{radar_equation}
P_{\mathrm{tar},k}=\frac{P_{\mathrm{t}}G_{\mathrm{Tx}}G_{\mathrm{Rx}}\sigma_k \lambda^2 L}{(4\pi)^3 d^4},
\end{equation}
where $P_{\mathrm{t}}$ is the transmit power, $G_{\mathrm{Tx}}$ and $G_{\mathrm{Rx}}$ are the \ac{Tx} and \ac{Rx} antenna gains, $\lambda$ is the wavelength, and $L$ collects unknown system losses/impairments. To eliminate hardware-dependent factors, we perform a dedicated free-space calibration by placing the \ac{Rx} at the target position (observation point) and measuring the received power $P_{\mathrm{Rx}}=\frac{P_{\mathrm{t}}G_{\mathrm{Tx}}G_{\mathrm{Rx}}\lambda^2 L}{(4\pi)^2 d^2}$, from which we compute the system factor, $K(\lambda,d)=\frac{P_{\mathrm{Rx}}}{4\pi d^2}$. The factor $4\pi d^2$ is multiplied in the denominator to match the powers of the single and double \ac{PL}. The calibrated \ac{RCS} estimate is then obtained snapshot-wise as $\sigma_k = K^{-1}(\lambda,d)\,P_{\mathrm{tar},k}$, ensuring that system losses and hardware impairments are removed. The calibration procedure compensates for deterministic system factors so that the inferred \ac{RCS} reflects the target scattering properties. Accordingly, we selected an operating distance that maximizes \ac{SCNR} while ensuring that relevant scatterers remain observable. With this constraint satisfied, the reported \ac{RCS} distributions are scalable to other operational ranges.}

\textcolor{black}{It is highlighted that whenever the motion/pose protocol produces repeated cycles, no cycle-wise averaging is performed. Instead, all snapshot-wise calibrated samples \(\{\sigma_k\}\) acquired over the entire measurement interval, i.e., pooled/concatenated across all cycles are aggregated into a single dataset for that \((f,\theta_b)\) configuration. Therefore, the resulting empirical \ac{PDF}/\ac{CDF} inherently captures the aspect/pose-dependent scattering realizations induced by the measurement protocol. In addition, since only a limited set of $\theta_b$ is sampled in the statistical campaign, we do not claim parameter invariance across continuous \(\theta_b\).}

All measurements are conducted under the measurement scope and domain of validity as delinted in Section \ref{sec2c}. Within this validity envelope, aspect diversity is obtained by executing a prescribed motion/pose protocol while recording discrete-time channel snapshots indexed by $k$. The $k$-th \ac{RCS} sample is associated with $\boldsymbol{\xi}_k=\boldsymbol{\xi}(t_k)$, where $t_k$ is the snapshot time, and the effective state increment between successive snapshots can be approximated by \(\Delta \boldsymbol{\xi}_k \approx \dot{\boldsymbol{\xi}}(t_k)\,\Delta t\) where \(\Delta t=t_{k+1}-t_k\) is the snapshot interval and \(\dot{\boldsymbol{\xi}}(t_k)\) is the time derivative of the aspect/pose state. Because the target orientation state \(\boldsymbol{\xi}(t)\) was not explicitly logged, we do not assign each snapshot to a fixed discrete angle grid. The reported results are therefore \textit{aspect-marginal (snapshot-amplitude) \ac{RCS}} statistics aggregated over the realized set \(\{\xi_k\}\), with the effective angular sampling governed by \(\Delta t\) and the executed motion. Extending the framework to aspect-conditioned statistics and temporal correlation/Doppler models, is left for future work. 

\textcolor{black}{It is important to clarify that the statistical \ac{RCS} distributions reported in this work are not intended to replace angle-dependent \ac{RCS} signatures. In the \ac{3GPP} \ac{RCS} modeling methodology, the target \ac{RCS} contains multiple components, including the mean or nominal \ac{RCS} level, the angular-dependent component, and a statistical variability component. The present measurements provide measurement-grounded target-class statistics for the mean \ac{RCS} level and the associated variability under the considered measurement protocol, frequency band, polarization, environment, and sensing geometries. 
Moreover, the measured \ac{RCS} variability is not due to noise but rather stems from the dynamic mechanical movements of the target itself (e.g., spinning \ac{UAV} rotors and chassis vibrations with continuous rotation of the \ac{UAV}) during the continuous measurement period, as well as the target's material properties and diffuse scattering. Finally, these statistical RCS fluctuations also account for the complex backscattering patterns of the targets. 
Therefore, the best-fit distributions, which are given in Tables \ref{mavic_25}-\ref{robot_28}, should be interpreted as aspect-/pose-marginal target-class \ac{RCS} statistics.
Consistent with 3GPP TR 38.901 \cite{3gpp38901}, an angularly-dependent \ac{RCS} model (RCS model 2 therein) is recommended alongside our derived statistical distributions for pose-dependent sensing tasks like posture recognition.}
\subsection{Parametric Distributions and Goodness-of-Fit Tests}\label{sec3b}
To model and analyze the \ac{RCS}, different parametric distributions such as normal, lognormal, gamma, and Weibull are systematically considered with each distribution tailored to capture specific statistical characteristics of the measured data. The normal distribution is suitable for datasets exhibiting symmetry around the mean. However, for positively skewed data, which is the case with the measured \ac{RCS}, the lognormal and Weibull distributions are more effective, as they can capture the asymmetry and heavy-tail behaviors associated with specular and diffused reflections from the targets. The gamma distribution is \textcolor{black}{also} well suited to model wide range of skewness and is therefore adept at characterizing \ac{RCS} data influenced by a mixture of strong and weak reflections.

Parametric distributions which best-fit the measured \ac{RCS} data enable the extraction of statistical parameters that not only characterize the data but also reveal reflective properties of the targets. The normal distribution is defined by the mean, $\mu$, and the standard deviation, $\sigma>0$. The lognormal distribution uses the mean, $\mu$, and the standard deviation, $\sigma\ge 0$, of the natural logarithm of the variable. The gamma distribution has a shape parameter, $A>0$, and a scale parameter, $B>0$, controlling the form and spread. The Weibull distribution is also defined by the shape $B>0$ and scale $A>0$. \textcolor{black}{We emphasize that the distribution families considered here are \textit{phenomenological surrogates} for the measured, aspect-/pose-marginal \ac{RCS} snapshots, rather than distributions tied uniquely to an target class. Different targets (and different motion/pose protocols) can yield different best-fit families because the instantaneous \ac{RCS} arises from different mixtures of (i) specular returns, (ii) diffuse scattering, and (iii) the effective number and relative strength of scattering centers that contribute over the sampled aspect/pose states. To this end, in case the observed power is the sum of many comparable scattering contributions, gamma/$\chi^2$-type behavior is physically plausible (Swerling-like composite targets). Conversely, when the return is dominated by intermittent alignment of a few strong facets, the snapshot ensemble becomes more right-skewed/heavy-tailed, for which lognormal/Weibull often provide best-fits. Since our statistics marginalize over the realized aspect/pose set $\{\xi_k\}$, the selected distribution parametrizes the \ac{RCS} induced by the target morphology/materials and the measurement protocol.}

\textcolor{black}{The \ac{GoF} of the parametric distributions is evaluated using the \ac{KS} statistic and \ac{MSE}. The \ac{GoF} matrices provide quantitative insight into the mismatch between the empirical \ac{CDF}, $F(x)$ and the fitted \ac{CDF}, $F_{\mathrm{fitted}}(x)$. \ac{KS} statistic measures the maximum absolute deviation between $F(x)$ and $F_{\mathrm{fitted}}(x)$ as:
\begin{equation}
\text{KS Statistic} = \max_x \bigl|F(x)-F_{\text{fitted}}(x)\bigr|.
\end{equation}
\ac{MSE} evaluates the average squared difference between $F(x_i)$ and $F_{\mathrm{fitted}}(x_i)$ across the discretized support $\{x_i\}_{i=1}^{N}$ as:
\begin{equation}
\text{MSE} = \frac{1}{N}\sum_{i=1}^{N}\Bigl(F(x_i)-F_{\mathrm{fitted}}(x_i)\Bigr)^2,
\end{equation}
where $N$ is the total number of \ac{CDF} evaluation points and $x_i$ are the bin edges.}

\textcolor{black}{It is worth noting that \ac{KS} and \ac{MSE} quantify different norms of the \ac{CDF} mismatch therefore, they may not necessarily nominate the same distribution as the best model for a given target and configuration. Specifically, \ac{KS} is an $\ell_\infty$-type criterion dominated by the \textit{worst-case local deviation}, whereas \ac{MSE} is an $\ell_2$-type criterion that emphasizes \textit{global average fidelity} across the full support.} \textcolor{black}{To this end,} our primary objective is to identify which of the considered parametric distributions best fits the empirical \ac{RCS} data. The \ac{GoF} metrics provide an understanding of fit quality by quantifying both localized (\ac{KS}) and global (\ac{MSE}) discrepancies between models and measurements. Smaller values of these metrics indicate a better fit, whereas elevated \ac{KS} and \ac{MSE} values indicate a loose fit. 
\subsection{RCS Fitting to Parametric Distributions}\label{sec3c}
This subsection quantifies the statistical behavior of the measured \ac{RCS} by fitting candidate parametric distributions to the empirical data. For each distribution, the associated \acp{PDF} and \acp{CDF} are matched to the \ac{PDF} and \ac{CDF} of the measured \ac{RCS} data. Owing to space limitations, we present the \acp{PDF} and \acp{CDF} for representative scenarios only, i.e., Mavic \(2\) Pro at \SI{25}{\GHz} under quasi-monostatic conditions, Matrice \(300\) RTK at \SI{25}{\GHz} with \(\theta_b \!=\! 20^\circ\), \ac{RA} at \SI{28}{\GHz} with \(\theta_b \!=\! 60^\circ\), and \ac{QR} at \SI{28}{\GHz} with \(\theta_b \!=\! 40^\circ\). The remaining \((f,\theta_b)\) combinations exhibit similar distributional trends. For completeness, the estimated fitting parameters for all targets and measurement configurations are reported in tabular form, enabling a comprehensive comparison of \ac{RCS} statistics across the dataset. Although \ac{KS} and \ac{MSE} rely on different \ac{GoF} criteria, they typically identify the same best-fitting distribution. \textcolor{black}{However, discrepancies occur only in a limited number of cases. Thus, the best fit distribution according to the \ac{KS} statistic is shaded in gray in the per target and configuration tables. If \ac{MSE} selects a different best fit distribution, it is explicitly indicated in black. When \ac{MSE} yields a tie between two distributions and one of them is also selected by \ac{KS}, that common distribution is adopted as the final best fit.}
\subsubsection{RCS Distribution Modeling for Unmanned Aerial Vehicles}
 \begin{figure}[tb]
    \centering
    \includegraphics[width=0.39\textwidth]{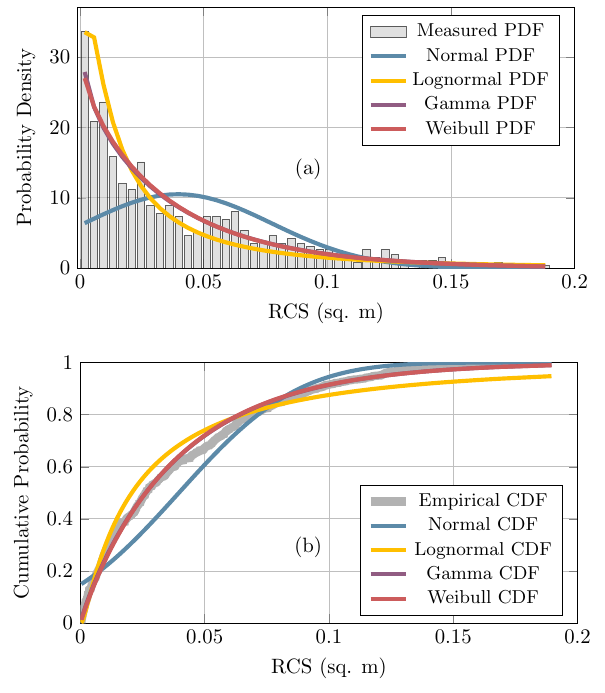} 
    \caption{RCS Distribution Modeling for Mavic \(2\) Pro at \(25\)GHz for \(\theta_b \approx 10^\circ\): (a) PDF Fitting, (b) CDF Fitting.}
    \label{mavic_25_0_results} 
\end{figure}
 \begin{figure}[h]
    \centering
    \includegraphics[width=0.39\textwidth]{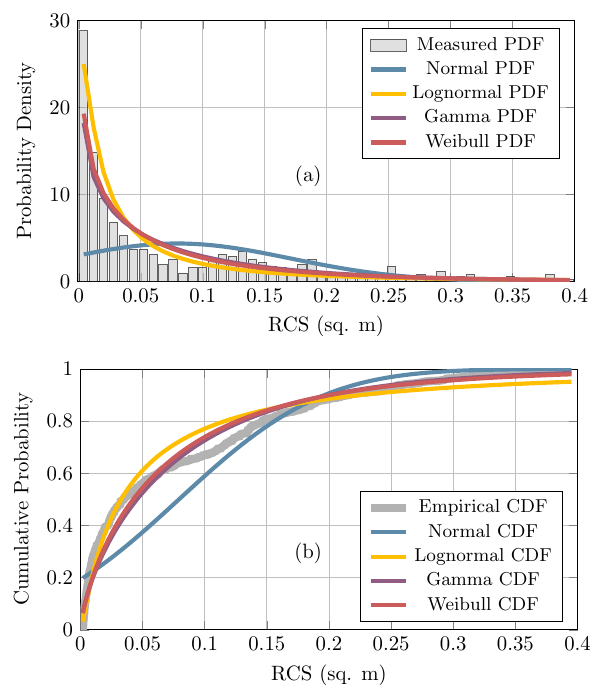} 
    \caption{RCS Distribution Modeling for Matrice \(300\) RTK at \(25\)GHz for \(\theta_b = 20^\circ\): (a) PDF Fitting, (b) CDF Fitting.}
    \label{rtk_25_20_results} 
\end{figure}

\textcolor{black}{Fig. \ref{mavic_25_0_results} depicts parametric distribution fitting to the measured \ac{RCS} of the Mavic \(2\) Pro at \SI{25}{\GHz} for \(\theta_b \approx 10^\circ\). The fitted \acp{PDF} and \acp{CDF} are shown in Fig. \ref{mavic_25_0_results}(a) and Fig. \ref{mavic_25_0_results}(b), respectively, while Table \ref{mavic_25} summarizes the \ac{GoF} across all \((f,\theta_b)\) configurations. From Fig. \ref{mavic_25_0_results}(a), we observe that the empirical \ac{RCS} exhibits a positive skew, for which heavy-tailed models such as lognormal, Weibull, and gamma provide the closest match, capturing both asymmetry and tail behavior. Overall backscatter levels are low, and the observed bimodality in the empirical \ac{PDF} is attributed to rotation-induced switching between dominant scattering mechanism, driven by abrupt changes in the contributions of asymmetric scatterers such as the camera gimbal. Consistent with the \ac{PDF} analysis, Fig. \ref{mavic_25_0_results}(b) shows that the gamma, Weibull, and lognormal \acp{CDF} also closely track the empirical \ac{CDF}, confirming their suitability for characterizing the observed \ac{RCS} statistics.}
\begin{table}[tb]
\centering
\small
\caption{Parameters at \SI{25}{\GHz} for varying \(\theta_b\) for Mavic.}
\begin{tabular}{|p{1.38cm}|p{1.05cm}|p{1.1cm}|p{3cm}|}
\hline
\multicolumn{4}{|c|}{\textbf{25 GHz, \(\theta_b \approx 10^\circ\)}} \\ \hline
\textbf{Dist.} & \textbf{KS Stat $(\times 10^{-2})$} & \textbf{MSE $(\times 10^{-3})$} & \textbf{Parameters} \\ \hline
Normal & \(15\) & \(6.7\) & \(\mu = 0.04, \sigma = 0.038\) \\
Lognormal & \(8.8\) & \(2.3\) & \(\mu = -3.9, \sigma = 1.4\) \\
\rowcolor{gray!20}
Gamma & \(5\) & \(0.43\) & \(A = 0.9, B = 0.044\) \\
Weibull & \(5.2\) & \(0.52\) & \(A = 0.039, B = 0.95\)
\\ \hline

\multicolumn{4}{|c|}{\textbf{25 GHz, \(\theta_b =20^\circ\)}} \\ \hline
Normal & \(16\) & \(7.1\) & \(\mu = 0.05, \sigma = 0.049\) \\
Lognormal & \(13\) & \(7.5\) & \(\mu = -3.90, \sigma = 1.67\) \\
\rowcolor{gray!20}
Gamma 	  & \(13\) & \(4.4\) & \(A = 0.66, B = 0.076\) \\
Weibull	 & \(14\) & \(5.1\) & \(A = 0.044, B = 0.77\) 
\\ \hline

\multicolumn{4}{|c|}{\textbf{25 GHz, \(\theta_b =40^\circ\)}} \\ \hline
Normal & \(18\) & \(11\) & \(\mu = 0.046, \sigma = 0.051\) \\
Lognormal & \(10\) & \(2.9\) & \(\mu = -3.92, \sigma = 1.64\) \\
\rowcolor{black!20}
Gamma & \(4.4\) & \(0.22\) & \(A = 0.70, B = 0.066\) \\
\rowcolor{gray!20}
Weibull & \(4.2\) & \(0.32\) & \(A = 0.041, B = 0.80\)
\\ \hline

\multicolumn{4}{|c|}{\textbf{25 GHz, \(\theta_b =60^\circ\)}} \\ \hline
Normal & \(16\) & \(6.3\) & \(\mu = 0.044, \sigma = 0.044\) \\
Lognormal & \(13\) & \(4.4\) & \(\mu = -3.96, \sigma = 1.64\) \\
\rowcolor{gray!20}
Gamma & \(8.1\) & \(1.6\) & \(A = 0.71, B = 0.062\) \\
Weibull & \(8.7\) & \(2\) & \(A = 0.04, B = 0.81\) 
\\ \hline
\end{tabular}
\label{mavic_25}
\end{table}
\begin{table}[tb]
\centering
\small
\caption{Parameters at \SI{26}{\GHz} for varying \(\theta_b\) for Mavic.}
\begin{tabular}{|p{1.38cm}|p{1.05cm}|p{1.1cm}|p{3.2cm}|}
\hline
\multicolumn{4}{|c|}{\textbf{26 GHz, \(\theta_b \approx 10^\circ\)}} \\ \hline
\textbf{Dist.} & \textbf{KS Stat $(\times 10^{-2})$} & \textbf{MSE $(\times 10^{-3})$} & \textbf{Parameters} \\ \hline
Normal & \(14\) & \(6.3\) & \(\mu = 0.024, \sigma = 0.014\) \\
\rowcolor{gray!20}
Lognormal & \(7.8\) & \(1.4\) & \(\mu = -3.8, \sigma = 0.52\) \\
Gamma & \(9.2\) & \(2.4\) & \(A = 3.66, B = 0.0066\) \\
Weibull & \(12\) & \(3\) & \(A = 0.027, B = 1.86\) 
\\ \hline

\multicolumn{4}{|c|}{\textbf{26 GHz, \(\theta_b =20^\circ\)}} \\ \hline
Normal & \(12\) & \(4\) & \(\mu = 0.036, \sigma = 0.031\) \\
Lognormal & \(12\) & \(5.4\) & \(\mu = -3.9, \sigma = 1.48\) \\
Gamma & \(7\) & \(1\) & \(A = 0.98, B = 0.036\) \\
\rowcolor{gray!20}
Weibull & \(6\) & \(0.83\) & \(A = 0.036, B = 1.02\) 
\\ \hline

\multicolumn{4}{|c|}{\textbf{26 GHz, \(\theta_b =40^\circ\)}} \\ \hline
Normal & \(34\) & \(42\) & \(\mu = 0.12, \sigma = 0.29\) \\
\rowcolor{gray!20}
Lognormal & \(5.4\) & \(1.1\) & \(\mu = -3.9, \sigma = 2.2\) \\
Gamma & \(15\) & \(6.8\) & \(A = 0.35, B = 0.35\) \\
Weibull & \(8.3\) & \(1.7\) & \(A = 0.056, B = 0.4996\) 
\\ \hline

\multicolumn{4}{|c|}{\textbf{26 GHz, \(\theta_b =60^\circ\)}} \\ \hline
Normal & \(29\) & \(26\) & \(\mu = 0.063, \sigma = 0.11\) \\
\rowcolor{gray!20}
Lognormal & \(3.1\) & \(0.2\) & \(\mu = -3.95, \sigma = 1.71\) \\
Gamma & \(9.4\) & \(2.8\) & \(A = 0.52, B = 0.12\) \\
Weibull & \(5.6\) & \(0.86\) & \(A = 0.043, B = 0.64\) 
\\ \hline

\end{tabular}
\label{mavic_26}
\end{table}
\begin{table}[h!]
\centering
\small
\caption{Parameters at \SI{27}{\GHz} for varying \(\theta_b\) for Mavic.}
\begin{tabular}{|p{1.38cm}|p{1.05cm}|p{1.1cm}|p{3cm}|}
\hline
\multicolumn{4}{|c|}{\textbf{27 GHz, \(\theta_b \approx 10^\circ\)}} \\ \hline
\textbf{Dist.} & \textbf{KS Stat}\newline $(\times 10^{-2})$ & \textbf{MSE}\newline$(\times 10^{-3})$ & \textbf{Parameters} \\ \hline
Normal & \(22\) & \(15\) & \(\mu = 0.067, \sigma = 0.088\) \\
\rowcolor{gray!20}
Lognormal & \(8.7\) & \(2.7\) & \(\mu = -3.83, \sigma = 1.74\) \\
Gamma & \(9.9\) & \(2.1\) & \(A = 0.55, B = 0.12\) \\
\rowcolor{black!20}
Weibull & \(9.4\) & \(1.8\) & \(A = 0.05, B = 0.66\) 
\\ \hline

\multicolumn{4}{|c|}{\textbf{27 GHz, \(\theta_b =20^\circ\)}} \\ \hline
Normal & \(16\) & \(8.4\) & \(\mu = 0.038, \sigma = 0.039\) \\
Lognormal & \(12\) & \(4.3\) & \(\mu = -3.9, \sigma = 1.46\) \\
Gamma & \(4.5\) & \(4\) & \(A = 0.89, B = 0.043\) \\
\rowcolor{gray!20}
Weibull & \(4.1\) & \(3.5\) & \(A = 0.038, B = 0.94\) 
\\ \hline

\multicolumn{4}{|c|}{\textbf{27 GHz, \(\theta_b =40^\circ\)}} \\ \hline
Normal & \(18\) & \(8.3\) & \(\mu = 0.053, \sigma = 0.06\) \\
Lognormal & \(11\) & \(4\) & \(\mu = -3.94, \sigma = 1.80\) \\
\rowcolor{gray!20}
Gamma & \(6.2\) & \(1.4\) & \(A = 0.6, B = 0.087\) \\
Weibull & \(7.6\) & \(1.9\) & \(A = 0.044, B = 0.72\) 
\\ \hline

\multicolumn{4}{|c|}{\textbf{27 GHz, \(\theta_b =60^\circ\)}} \\ \hline
Normal & \(25\) & \(21\) & \(\mu = 0.038, \sigma = 0.049\) \\
\rowcolor{gray!20}
Lognormal & \(9.3\) & \(3.3\) & \(\mu = -3.97, \sigma = 1.17\) \\
Gamma & \(15\) & \(8.2\) & \(A = 0.82, B = 0.046\) \\
Weibull & \(13\) & \(6\) & \(A = 0.034, B = 0.84\) 
\\ \hline

\end{tabular}
\label{mavic_27}
\end{table}
\begin{table}[h!]
\centering
\small
\caption{Parameters at  \SI{28}{\GHz} for varying \(\theta_b\) for Mavic.}
\begin{tabular}{|p{1.38cm}|p{1.05cm}|p{1.1cm}|p{3cm}|}
\hline
\multicolumn{4}{|c|}{\textbf{28 GHz, \(\theta_b \approx 10^\circ\)}} \\ \hline
\textbf{Dist.} & \textbf{KS Stat}\newline$(\times 10^{-2})$ & \textbf{MSE}\newline$(\times 10^{-3})$ & \textbf{Parameters} \\ \hline
Normal & \(19\) & \(11\) & \(\mu = 0.027, \sigma = 0.023\) \\
\rowcolor{gray!20}
Lognormal & \(6\) & \(0.81\) & \(\mu = -3.79, \sigma = 0.61\) \\
Gamma & \(8.4\) & \(2.4\) & \(A = 2.50, B = 0.011\) \\
 Weibull & \(13\) & \(3.8\) & \(A = 0.031, B = 1.42\)
\\ \hline

\multicolumn{4}{|c|}{\textbf{28 GHz, \(\theta_b =20^\circ\)}} \\ \hline
Normal & \(12\) & \(4.6\) & \(\mu = 0.045, \sigma = 0.04\) \\
Lognormal & \(15\) & \(8.6\) & \(\mu = -3.81, \sigma = 1.77\) \\
Gamma & \(8.6\) & \(1.8\) & \(A = 0.80, B = 0.056\) \\
\rowcolor{gray!20}
Weibull & \(7.5\) & \(1.3\) & \(A = 0.044, B = 0.92\) 
\\ \hline

\multicolumn{4}{|c|}{\textbf{28 GHz, \(\theta_b =40^\circ\)}} \\ \hline
Normal & \(25\) & \(21\) & \(\mu = 0.027, \sigma = 0.029\) \\
\rowcolor{gray!20}
Lognormal & \(9.2\) & \(1.9\) & \(\mu = -3.88, \sigma = 0.69\) \\
Gamma & \(13\) & \(6.1\) & \(A = 1.83, B = 0.015\) \\
Weibull & \(18\) & \(7.3\) & \(A = 0.029, B = 1.21\) 
\\ \hline

\multicolumn{4}{|c|}{\textbf{28 GHz, \(\theta_b =60^\circ\)}} \\ \hline
Normal & \(20\) & \(15\) & \(\mu = 0.04, \sigma = 0.047\) \\
\rowcolor{gray!20}
Lognormal & \(4.5\) & \(0.57\) & \(\mu = -3.91, \sigma = 1.24\) \\
Gamma & \(8.5\) & \(2.4\) & \(A = 0.84, B = 0.047\) \\
Weibull & \(7.5\) & \(1.5\) & \(A = 0.037, B = 0.87\) 
\\ \hline
\end{tabular}
\label{mavic_28}
\end{table}

\textcolor{black}{Fig. \ref{rtk_25_20_results} presents the parametric distribution fitting for the measured \ac{RCS} of the Matrice \(300\) RTK at \SI{25}{\GHz} with \(\theta_b=20^\circ\). Owing to its larger physical dimensions relative to the Mavic \(2\) Pro, the Matrice \(300\) RTK is expected to yield a higher \ac{RCS}, primarily due to the increased effective reflecting area resulting in stronger backscattered power. In addition, the rear-mounted lithium-ion battery packs acts as a prominent specular scatterer and is anticipated to dominate the overall scattering response, thereby elevating the measured \ac{RCS}. Again, the bimodal structure observed in the empirical \ac{RCS} distribution is consistent with rotation-dependent intermittency of these battery-induced specular returns.}

\textcolor{black}{Fig. \ref{rtk_25_20_results}(a) reveals that the \ac{RCS} distribution for Matrice \(300\) RTK also exhibits positive skewness, characterized by a dominance of low-magnitude reflections and occasional high-magnitude specular reflections. However, the specular reflections are notably more pronounced compared to the Mavic \(2\) Pro (cf. Fig. \ref{mavic_25_0_results}(a)). Statistical models, including the lognormal, Weibull, and gamma distributions, effectively capture the underlying \ac{RCS} characteristics as evident from the fitted \acp{PDF}, as summarized in Table \ref{rtk_25}. The empirical \ac{CDF} fitting further supports this observation in Fig. \ref{rtk_25_20_results}(b), where the Weibull, lognormal, and gamma distributions exhibit strong consistency with the empirical \ac{CDF}, where these three distributions closely track the empirical \ac{CDF}.}

\textcolor{black}{The parametric fitting results for the measured \ac{RCS} of the Mavic \(2\) Pro across the considered \((\theta_b,f)\) configurations are summarized in Tables~\ref{mavic_25}-\ref{mavic_28}. The corresponding results for the Matrice \(300\) RTK, evaluated over the same set of (\(\theta_b,f\)) are reported in Tables \ref{rtk_25}-\ref{rtk_28}. \ac{GoF} is quantified using both the \ac{KS} statistic and the \ac{MSE}. From the results in Tables \ref{mavic_25}-\ref{mavic_28} and Tables \ref{rtk_25}-\ref{rtk_28}, following observations are drawn:}
\begin{enumerate}
    \item The Weibull, lognormal, and gamma distributions consistently provide the best fits. Specifically, based on \ac{KS} \ac{GoF}, the gamma distribution demonstrates superior performance in \(12\) out of the \(32\) representative cases, encompassing various (\(\theta_b,f\)) combinations. In contrast, the lognormal and Weibull distributions achieve the best fit in \(9\) and \(11\) cases, respectively. However, we can also observe that the differences in the \ac{GoF} metrics among these distributions are minimal in most representative cases. Thus, any of these three distributions can effectively model the \ac{RCS} of the \acp{UAV}. 
    \item Considering the parameters of the best fit distributions, \textcolor{black}{we observe from Tables \ref{mavic_25}-\ref{rtk_28}} that an increase in frequency for a given \(\theta_b\) marginally increases the \ac{RCS}. \textcolor{black}{The marginal increase in frequency is evident in Fig.~\ref{rcs_thetab_fc}(a) for the Mavic \(2\) Pro, where the \(\mu\) parameter derived from lognormal distribution generally increases with increasing frequency.}
    \item For both \acp{UAV}, Tables \ref{mavic_25}-\ref{rtk_28} indicate a systematic reduction in \ac{RCS} with increasing \(\theta_b\) across all measured frequencies. \textcolor{black}{The reduction in \ac{RCS} is also visible for Mavic \(2\) Pro in Fig.~\ref{rcs_thetab_fc}(a), where \(\mu\) parameter of the lognormal distribution decreases with an increase in \(\theta_b\).} 
    \item \textcolor{black}{From the estimated parameters of the selected best fit models, Matrice \(300\) RTK consistently exhibits higher \ac{RCS} than the Mavic \(2\) Pro, which is primarily attributed to its larger physical size and the lithium-ion batteries, which intermittently generate strong specular returns. Although material properties also affect scattering, their isolated impact on \ac{RCS} is difficult to quantify.}
\end{enumerate}
\begin{table}[tb]
\centering
\small
\caption{Parameters at \SI{25}{\GHz} for varying \(\theta_b\) for Matrice.}
\begin{tabular}{|p{1.38cm}|p{1.05cm}|p{1.1cm}|p{3cm}|}
\hline
\multicolumn{4}{|c|}{\textbf{25 GHz, \(\theta_b \approx 10^\circ\)}} \\ \hline
\textbf{Dist.} & \textbf{KS Stat}\newline$(\times 10^{-2})$ & \textbf{MSE}\newline$(\times 10^{-3})$ & \textbf{Parameters} \\ \hline
Normal & \(19\) & \(11\) & \(\mu = 0.063, \sigma = 0.072\) \\
Lognormal & \(10\) & \(2.3\) & \(\mu = -3.5, \sigma = 1.42\) \\
Gamma & \(4.1\) & \(0.4\) & \(A = 0.79, B = 0.079\) \\
\rowcolor{gray!20}
Weibull & \(4\) & \(0.4\) & \(A = 0.058, B = 0.86\) 
\\ \hline

\multicolumn{4}{|c|}{\textbf{25 GHz, \(\theta_b =20^\circ\)}} \\ \hline
Normal & \(20\) & \(13\) & \(\mu = 0.075, \sigma = 0.09\) \\
Lognormal & \(10\) & \(2.2\) & \(\mu = -3.54, \sigma = 1.59\) \\
Gamma & \(8.7\) & \(2.6\) & \(A = 0.63, B = 0.12\) \\
\rowcolor{gray!20}
Weibull & \(8.1\) & \(2.2\) & \(A = 0.062, B = 0.73\) 
\\ \hline

\multicolumn{4}{|c|}{\textbf{25 GHz, \(\theta_b =40^\circ\)}} \\ \hline
Normal & \(19\) & \(12\) & \(\mu = 0.054, \sigma = 0.063\) \\
Lognormal & \(8.6\) & \(2.4\) & \(\mu = -3.59, \sigma = 1.36\) \\
Gamma & \(4.6\) & \(0.32\) & \(A = 0.85, B = 0.063\) \\
\rowcolor{gray!20}
Weibull & \(4.2\) & \(0.29\) & \(A = 0.051, B = 0.89\) 
\\ \hline

\multicolumn{4}{|c|}{\textbf{25 GHz, \(\theta_b =60^\circ\)}} \\ \hline
Normal & \(20\) & \(13\) & \(\mu = 0.072, \sigma = 0.089\) \\
Lognormal & \(8\) & \(1.7\) & \(\mu = -3.62, \sigma = 1.70\) \\
\rowcolor{gray!20}
Gamma & \(4.6\) & \(0.57\) & \(A = 0.61, B = 0.11\) \\
\rowcolor{black!20}
Weibull & \(4.9\) & \(0.53\) & \(A = 0.059, B = 0.72\) 
\\ \hline

\end{tabular}
\label{rtk_25}
\end{table}
\begin{table}[tb]
\centering
\small
\caption{Parameters at \SI{26}{\GHz} for varying \(\theta_b\) for Matrice.}
\begin{tabular}{|p{1.38cm}|p{1.05cm}|p{1.1cm}|p{3.2cm}|}
\hline
\multicolumn{4}{|c|}{\textbf{26 GHz, \(\theta_b \approx 10^\circ\)}} \\ \hline
\textbf{Dist.} & \textbf{KS Stat}\newline$(\times 10^{-2})$ & \textbf{MSE}\newline$(\times 10^{-3})$ & \textbf{Parameters} \\ \hline
Normal & \(19\) & \(11\) & \(\mu = 0.066, \sigma = 0.074\) \\
Lognormal & \(7.1\) & \(1.3\) & \(\mu = -3.49, \sigma = 1.47\) \\
Gamma & \(3.3\) & \(0.26\) & \(A = 0.77, B = 0.085\) \\
\rowcolor{gray!20}
Weibull & \(2.2\) & \(1.3\) & \(A = 0.06, B = 0.84\) 
\\ \hline

\multicolumn{4}{|c|}{\textbf{26 GHz, \(\theta_b =20^\circ\)}} \\ \hline
Normal & \(17\) & \(8.6\) & \(\mu = 0.097, \sigma = 0.1\) \\
Lognormal & \(15\) & \(7.5\) & \(\mu = -3.53, \sigma = 2.26\) \\
\rowcolor{gray!20}
Gamma & \(7.6\) & \(1.6\) & \(A = 0.52, B = 0.18\) \\
Weibull & \(8.4\) & \(2.3\) & \(A = 0.077, B = 0.65\)
\\ \hline

\multicolumn{4}{|c|}{\textbf{26 GHz, \(\theta_b =40^\circ\)}} \\ \hline
Normal & \(17\) & \(7.2\) & \(\mu = 0.08, \sigma = 0.085\) \\
Lognormal & \(14\) & \(6.5\) & \(\mu = -3.56, \sigma = 2\) \\
\rowcolor{gray!20}
Gamma & \(8.1\) & \(1.5\) & \(A = 0.59, B = 0.13\) \\
Weibull & \(8.9\) & \(2\) & \(A = 0.06, B = 0.72\) 
\\ \hline

\multicolumn{4}{|c|}{\textbf{26 GHz, \(\theta_b =60^\circ\)}} \\ \hline
Normal & \(20\) & \(13\) & \(\mu = 0.074, \sigma = 0.089\) \\
Lognormal & \(8.6\) & \(2.3\) & \(\mu = -3.59, \sigma = 1.7462\) \\
\rowcolor{gray!20}
Gamma & \(3.4\) & \(0.29\) & \(A = 0.61, B = 0.12\) \\
Weibull & \(4.3\) & \(0.34\) & \(A = 0.061, B = 0.72\) 
\\ \hline
\end{tabular}
\label{rtk_26}
\end{table}
\begin{table}[h!]
\centering
\small
\caption{Parameters at \SI{27}{\GHz} for varying \(\theta_b\) for Matrice.}
\begin{tabular}{|p{1.38cm}|p{1.05cm}|p{1.1cm}|p{3cm}|}
\hline
\multicolumn{4}{|c|}{\textbf{27 GHz, \(\theta_b \approx 10^\circ\)}} \\ \hline
\textbf{Dist.} & \textbf{KS Stat}\newline$(\times 10^{-2})$ & \textbf{MSE}\newline$(\times 10^{-3})$ & \textbf{Parameters} \\ \hline
Normal & \(16\) & \(7.2\) & \(\mu = 0.092, \sigma = 0.093\) \\
Lognormal & \(14\) & \(6\) & \(\mu = -3.47, \sigma = 1.96\) \\
\rowcolor{gray!20}
Gamma & \(9.3\) & \(2.9\) & \(A = 0.57, B = 0.16\) \\
Weibull & \(10\) & \(3.7\) & \(A = 0.075, B = 0.69\) 
\\ \hline

\multicolumn{4}{|c|}{\textbf{27 GHz, \(\theta_b =20^\circ\)}} \\ \hline
Normal & \(20\) & \(12\) & \(\mu = 0.077, \sigma = 0.092\) \\
Lognormal & \(7.8\) & \(1.4\) & \(\mu = -3.49, \sigma = 1.63\) \\
Gamma & \(5.7\) & \(0.78\) & \(A = 0.64, B = 0.11\) \\
\rowcolor{gray!20}
Weibull & \(4.6\) & \(0.67\) & \(A = 0.065, B = 0.74\) 
\\ \hline

\multicolumn{4}{|c|}{\textbf{27 GHz, \(\theta_b =40^\circ\)}} \\ \hline
Normal & \(22\) & \(18\) & \(\mu = 0.057, \sigma = 0.071\) \\
\rowcolor{gray!20}
Lognormal & \(3.9\) & \(0.4\) & \(\mu = -3.53, \sigma = 1.19\) \\
Gamma & \(10\) & \(3.5\) & \(A = 0.86, B = 0.065\) \\
Weibull & \(8.6\) & \(2.2\) & \(A = 0.053, B = 0.87\) 
\\ \hline

\multicolumn{4}{|c|}{\textbf{27 GHz, \(\theta_b =60^\circ\)}} \\ \hline
Normal & \(18\) & \(10\) & \(\mu = 0.078, \sigma = 0.089\) \\
Lognormal & \(10\) & \(4\) & \(\mu = -3.58, \sigma = 1.87\) \\
\rowcolor{gray!20}
Gamma & \(5.2\) & \(0.5\) & \(A = 0.59, B = 0.13\) \\
Weibull & \(6.3\) & \(0.96\) & \(A = 0.064, B = 0.71\) 
\\ \hline
\end{tabular}
\label{rtk_27}
\end{table}
\begin{table}[h!]
\centering
\small
\caption{Parameters at \SI{28}{\GHz} for varying \(\theta_b\) for Matrice.}
\begin{tabular}{|p{1.38cm}|p{1.05cm}|p{1.1cm}|p{3cm}|}
\hline
\multicolumn{4}{|c|}{\textbf{28 GHz, \(\theta_b \approx 10^\circ\)}} \\ \hline
\textbf{Dist.} & \textbf{KS Stat}\newline$(\times 10^{-2})$ & \textbf{MSE}\newline$(\times 10^{-3})$ & \textbf{Parameters} \\ \hline
Normal & \(18\) & \(10\) & \(\mu = 0.066, \sigma = 0.073\) \\
Lognormal & \(8.2\) & \(1.6\) & \(\mu = -3.49, \sigma = 1.48\) \\
\rowcolor{black!20}
Gamma & \(4.7\) & \(0.41\) & \(A = 0.76, B = 0.087\) \\
\rowcolor{gray!20}
Weibull & \(4.3\) & \(0.42\) & \(A = 0.06, B = 0.83\) 
\\ \hline

\multicolumn{4}{|c|}{\textbf{28 GHz, \(\theta_b =20^\circ\)}} \\ \hline
Normal & \(20\) & \(13\) & \(\mu = 0.081, \sigma = 0.097\) \\
Lognormal & \(8.6\) & \(1.8\) & \(\mu = -3.52, \sigma = 1.67\) \\
Gamma & \(7.9\) & \(1.5\) & \(A = 0.607, B = 0.13\) \\
\rowcolor{gray!20}
Weibull & \(7\) & \(1.3\) & \(A = 0.065, B = 0.71\) 
\\ \hline

\multicolumn{4}{|c|}{\textbf{28 GHz, \(\theta_b =40^\circ\)}} \\ \hline
Normal & \(19\) & \(12\) & \(\mu = 0.084, \sigma = 0.099\) \\
Lognormal & \(8.5\) & \(2.3\) & \(\mu = -3.56, \sigma = 1.84\) \\
\rowcolor{gray!20}
Gamma & \(5.2\) & \(0.69\) & \(A = 0.57, B = 0.14\) \\
Weibull & \(5.2\) & \(0.87\) & \(A = 0.067, B = 0.69\) 
\\ \hline

\multicolumn{4}{|c|}{\textbf{28 GHz, \(\theta_b =60^\circ\)}} \\ \hline
Normal & \(19\) & \(11\) & \(\mu = 0.083, \sigma = 0.095\) \\
Lognormal & \(10\) & \(4.9\) & \(\mu = -3.57, \sigma = 2.14\) \\
\rowcolor{gray!20}
Gamma & \(3.2\) & \(0.26\) & \(A = 0.57, B = 0.14\) \\
Weibull & \(3.9\) & \(0.43\) & \(A = 0.068, B = 0.7\) 
\\ \hline
\end{tabular}
\label{rtk_28}
\end{table}
\subsubsection{RCS Distribution Modeling for Robotic Arm}
 \begin{figure}[t]
    \centering
    \includegraphics[width=0.4\textwidth]{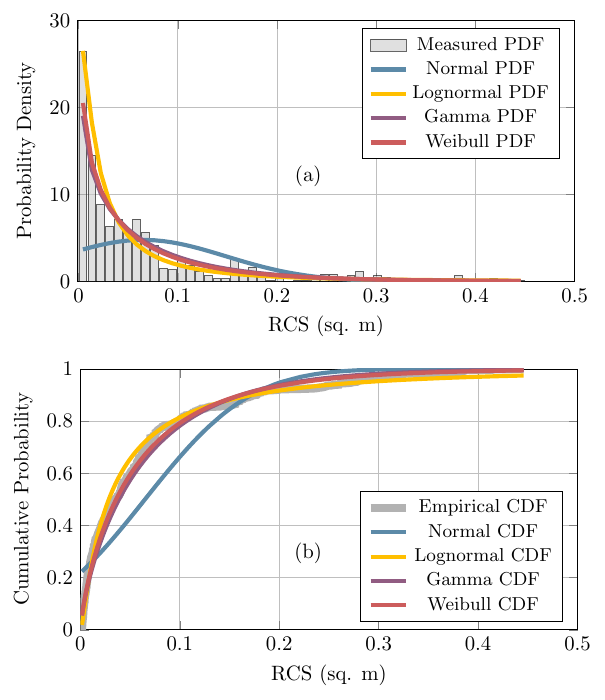} 
    \caption{RCS Distribution Modeling for RA at \(28\)GHz for \(\theta_b = 60^\circ\): (a) PDF Fitting, (b) CDF Fitting.}
    \label{ra_28_60_results} 
\end{figure}

\textcolor{black}{Fig. \ref{ra_28_60_results} shows \acp{RA}' empirical \ac{RCS} data at \SI{28}{\GHz} for of $\theta_b = 60^\circ$ and compares several parametric distribution fitting. The \ac{PDF} in Fig. \ref{ra_28_60_results}(a) shows that the Weibull, gamma, and lognormal distributions closely match the measured data. In contrast, the normal distribution does not represent the measured data well, as it fails to capture the large number of weak reflections and the small number of strong specular reflections, highlighting its limitation in modeling asymmetric data. The empirical \ac{CDF} as illustrated in Fig. \ref{ra_28_60_results}(b) further supports these observations.}

\textcolor{black}{The fitting results of the measured \ac{RCS} data of the \ac{RA} for different parametric distributions are summarized in Tables \ref{ra_25}-\ref{ra_28}. Same \ac{GoF} metrics used for the \acp{UAV} are applied to evaluate the accuracy of each distribution. Based on the results in Tables \ref{ra_25}-\ref{ra_28}, following observations are made:}
\begin{enumerate}
    \item \textcolor{black}{The lognormal, Weibull, and gamma distributions provide the best fit for the measured \ac{RCS} data of the \ac{RA}. Among these, the gamma and Weibull distributions perform best in $8$ and $7$ out of the $16$ representative cases, respectively, across different $(\theta_b,f)$ combinations based on \ac{KS} \ac{GoF}. The lognormal distribution yields the best fit in the remaining $2$ cases. Similar to the results observed for \acp{UAV}, the differences in the \ac{GoF} metrics among these three distributions are small in most cases.}
   \item \textcolor{black}{The estimated parameters of the best-fitting distributions indicate that increasing $\theta_b$ results in a decrease in the \ac{RCS} at a fixed frequency, which is consistent with the trends previously observed for \acp{UAV}.}
    \item \textcolor{black}{The \ac{RCS} is also observed to decrease with increasing frequency for a fixed $\theta_b$, in agreement with the behavior reported for \acp{UAV}.}
    \item \textcolor{black}{The \ac{RCS} of the \ac{RA} is significantly higher than that of \acp{UAV}. The difference is mainly due to the material composition of the \ac{RA}, which includes materials that produce stronger backscatter, whereas \acp{UAV} are primarily made of materials such as plastic and carbon fiber with lower reflectivity, leading to reduced \ac{RCS} values.}
\end{enumerate}
\begin{table}[t]
\centering
\small
\caption{Parameters at \SI{25}{\GHz} for varying \(\theta_b\) for RA.}
\begin{tabular}{|p{1.38cm}|p{1.05cm}|p{1.1cm}|p{3cm}|}
\hline
\multicolumn{4}{|c|}{\textbf{25 GHz, \(\theta_b \approx 10^\circ\)}} \\ \hline
\textbf{Dist.} & \textbf{KS Stat}\newline$(\times 10^{-2})$ & \textbf{MSE}\newline$(\times 10^{-3})$ & \textbf{Parameters} \\ \hline
Normal & \(20\) & \(13\) & \(\mu = 0.085, \sigma = 0.098\) \\
Lognormal & \(11\) & \(3.2\) & \(\mu = -3.43, \sigma = 1.79\) \\
\rowcolor{gray!20}
Gamma & \(4.3\) & \(0.34\) & \(A = 0.63, B = 0.13\) \\
Weibull & \(4.5\) & \(0.45\) & \(A = 0.072, B = 0.74\) 
\\ \hline

\multicolumn{4}{|c|}{\textbf{25 GHz, \(\theta_b =20^\circ\)}} \\ \hline
Normal & \(18\) & \(9.4\) & \(\mu = 0.10, \sigma = 0.11\) \\
Lognormal & \(11\) & \(4.7\) & \(\mu = -3.48, \sigma = 2.23\) \\
\rowcolor{gray!20}
Gamma & \(8.4\) & \(1.1\) & \(A = 0.52, B = 0.19\) \\
Weibull & \(9.7\) & \(1.6\) & \(A = 0.08, B = 0.65\) 
\\ \hline

\multicolumn{4}{|c|}{\textbf{25 GHz, \(\theta_b =40^\circ\)}} \\ \hline
Normal & \(11\) & \(3.3\) & \(\mu = 0.04, \sigma = 0.03\) \\
Lognormal & \(10\) & \(3.1\) & \(\mu = -3.5, \sigma = 1.08\) \\
Gamma & \(7.1\) & \(1.1\) & \(A = 1.29, B = 0.03\) \\
\rowcolor{gray!20}
Weibull & \(6.3\) & \(1.1\) & \(A = 0.049, B = 1.21\) 
\\ \hline

\multicolumn{4}{|c|}{\textbf{25 GHz, \(\theta_b =60^\circ\)}} \\ \hline
Normal & \(16\) & \(7.5\) & \(\mu = 0.09, \sigma = 0.09\) \\
Lognormal & \(15\) & \(9.5\) & \(\mu = -3.52, \sigma = 2.35\) \\
\rowcolor{gray!20}
Gamma & \(8\) & \(1.9\) & \(A = 0.55, B = 0.16\) \\
Weibull & \(8.4\) & \(2\) & \(A = 0.076, B = 0.69\) 
\\ \hline

\end{tabular}
\label{ra_25}
\end{table}
\begin{table}[h!]
\centering
\small
\caption{Parameters at \SI{25}{\GHz} for varying \(\theta_b\) for RA.}
\begin{tabular}{|p{1.38cm}|p{1.05cm}|p{1.1cm}|p{3cm}|}
\hline
\multicolumn{4}{|c|}{\textbf{26 GHz, \(\theta_b \approx 10^\circ\)}} \\ \hline
\textbf{Dist.} & \textbf{KS Stat}\newline$(\times 10^{-2})$ & \textbf{MSE}\newline$(\times 10^{-3})$ & \textbf{Parameters} \\ \hline
Normal & \(20\) & \(13\) & \(\mu = 0.08, \sigma = 0.09\) \\
Lognormal & \(11\) & \(3.4\) & \(\mu = -3.45, \sigma = 1.83\) \\
\rowcolor{gray!20}
Gamma & \(4.3\) & \(0.34\) & \(A = 0.62, B = 0.13\) \\
Weibull & \(4.4\) & \(0.46\) & \(A = 0.07, B = 0.73\) 
\\ \hline

\multicolumn{4}{|c|}{\textbf{26 GHz, \(\theta_b =20^\circ\)}} \\ \hline
Normal & \(15\) & \(6.2\) & \(\mu = 0.058, \sigma = 0.058\) \\
Lognormal & \(10\) & \(3.1\) & \(\mu = -3.49, \sigma = 1.41\) \\
\rowcolor{gray!20}
Gamma & \(5.3\) & \(0.51\) & \(A = 0.88, B = 0.066\) \\
Weibull & \(5.4\) & \(0.57\) & \(A = 0.057, B = 0.93\) 
\\ \hline

\multicolumn{4}{|c|}{\textbf{26 GHz, \(\theta_b =40^\circ\)}} \\ \hline
Normal & \(21\) & \(15\) & \(\mu = 0.08, \sigma = 0.1\) \\
Lognormal & \(10\) & \(1.8\) & \(\mu = -3.54, \sigma = 1.64\) \\
Gamma & \(8.9\) & \(2.3\) & \(A = 0.59, B = 0.13\) \\
\rowcolor{gray!20}
Weibull & \(7.5\) & \(1.8\) & \(A = 0.064, B = 0.7\) 
\\ \hline

\multicolumn{4}{|c|}{\textbf{26 GHz, \(\theta_b =60^\circ\)}} \\ \hline
Normal & \(25\) & \(20\) & \(\mu = 0.069, \sigma = 0.089\) \\
\rowcolor{gray!20}
Lognormal & \(5.1\) & \(0.62\) & \(\mu = -3.55, \sigma = 1.47\) \\
Gamma & \(11\) & \(3.8\) & \(A = 0.68, B = 0.1\) \\
Weibull & \(9.7\) & \(2.4\) & \(A = 0.058, B = 0.76\) 
\\ \hline
\end{tabular}
\label{ra_26}
\end{table}
\begin{table}[h!]
\centering
\small
\caption{Parameters at \SI{27}{\GHz} for varying \(\theta_b\) for RA.}
\begin{tabular}{|p{1.38cm}|p{1.05cm}|p{1.1cm}|p{3cm}|}
\hline
\multicolumn{4}{|c|}{\textbf{27 GHz, \(\theta_b \approx 10^\circ\)}} \\ \hline
\textbf{Dist.} & \textbf{KS Stat}\newline$(\times 10^{-2})$ & \textbf{MSE}\newline$(\times 10^{-3})$ & \textbf{Parameters} \\ \hline
Normal & \(20\) & \(12\) & \(\mu = 0.089, \sigma = 0.1\) \\
Lognormal & \(8.5\) & \(2.6\) & \(\mu = -3.48, \sigma = 1.82\) \\
\rowcolor{gray!20}
Gamma & \(6.9\) & \(0.96\) & \(A = 0.5820, B = 0.15\) \\
Weibull & \(7.1\) & \(1.1\) & \(A = 0.072, B = 0.69\) 
\\ \hline

\multicolumn{4}{|c|}{\textbf{27 GHz, \(\theta_b =20^\circ\)}} \\ \hline
Normal & \(21\) & \(13\) & \(\mu = 0.083, \sigma = 0.10\) \\
Lognormal & \(9.2\) & \(2\) & \(\mu = -3.48, \sigma = 1.75\) \\
Gamma & \(4.3\) & \(0.32\) & \(A = 0.61, B = 0.13\) \\
\rowcolor{gray!20}
Weibull & \(3.4\) & \(0.3\) & \(A = 0.068, B = 0.72\) 
\\ \hline

\multicolumn{4}{|c|}{\textbf{27 GHz, \(\theta_b =40^\circ\)}} \\ \hline
Normal & \(13\) & \(7.4\) & \(\mu = 0.032, \sigma = 0.015\) \\
\rowcolor{gray!20}
Lognormal & \(6.1\) & \(0.6\) & \(\mu = -3.52, \sigma = 0.38\) \\
Gamma & \(7.5\) & \(1.8\) & \(A = 6.18, B = 0.0052\) \\
Weibull & \(14\) & \(6\) & \(A = 0.036, B = 2.17\) 
\\ \hline

\multicolumn{4}{|c|}{\textbf{27 GHz, \(\theta_b =60^\circ\)}} \\ \hline
Normal & \(21\) & \(14\) & \(\mu = 0.074, \sigma = 0.089\) \\
Lognormal & \(6.1\) & \(0.94\) & \(\mu = -3.54, \sigma = 1.59\) \\
Gamma & \(6.9\) & \(1.3\) & \(A = 0.64, B = 0.11\) \\
\rowcolor{gray!20}
Weibull & \(6.1\) & \(0.89\) & \(A = 0.062, B = 0.74\) 
\\ \hline
\end{tabular}
\label{ra_27}
\end{table}
\begin{table}[h!]
\centering
\small
\caption{Parameters at \SI{28}{\GHz} for varying \(\theta_b\) for RA.}
\begin{tabular}{|p{1.38cm}|p{1.05cm}|p{1.1cm}|p{3cm}|}
\hline
\multicolumn{4}{|c|}{\textbf{28 GHz, \(\theta_b \approx 10^\circ\)}} \\ \hline
\textbf{Dist.} & \textbf{KS Stat}\newline$(\times 10^{-2})$ & \textbf{MSE}\newline$(\times 10^{-3})$ & \textbf{Parameters} \\ \hline
Normal & \(15\) & \(7\) & \(\mu = 0.069, \sigma = 0.069\) \\
Lognormal & \(13\) & \(5\) & \(\mu = -3.49, \sigma = 1.67\) \\
\rowcolor{gray!20}
Gamma & \(5.8\) & \(0.78\) & \(A = 0.73, B = 0.094\) \\
Weibull & \(5.8\) & \(0.99\) & \(A = 0.063, B = 0.83\) 
\\ \hline

\multicolumn{4}{|c|}{\textbf{28 GHz, \(\theta_b =20^\circ\)}} \\ \hline
Normal & \(19\) & \(11\) & \(\mu = 0.062, \sigma = 0.072\) \\
Lognormal & \(8.5\) & \(1.7\) & \(\mu = -3.52, \sigma = 1.39\) \\
Gamma & \(6.3\) & \(0.68\) & \(A = 0.78, B = 0.079\) \\
\rowcolor{gray!20}
Weibull & \(5.7\) & \(0.56\) & \(A = 0.057, B = 0.84\) 
\\ \hline

\multicolumn{4}{|c|}{\textbf{28 GHz, \(\theta_b =40^\circ\)}} \\ \hline
Normal & \(17\) & \(8.8\) & \(\mu = 0.09, \sigma = 0.095\) \\
Lognormal & \(13\) & \(6\) & \(\mu = -3.54, \sigma = 2.24\) \\
\rowcolor{gray!20}
Gamma & \(7.7\) & \(1.2\) & \(A = 0.54, B = 0.16\) \\
Weibull & \(8.7\) & \(1.7\) & \(A = 0.07, B = 0.68\) 
\\ \hline

\multicolumn{4}{|c|}{\textbf{28 GHz, \(\theta_b =60^\circ\)}} \\ \hline
Normal & \(22\) & \(16\) & \(\mu = 0.064, \sigma = 0.083\) \\
Lognormal & \(7\) & \(1\) & \(\mu = -3.55, \sigma = 1.40\) \\
Gamma & \(6.5\) & \(1.3\) & \(A = 0.73, B = 0.08\) \\
\rowcolor{gray!20}
Weibull & \(5.4\) & \(0.67\) & \(A = 0.056, B = 0.79\) 
\\ \hline
\end{tabular}
\label{ra_28}
\end{table}
\subsubsection{RCS Distribution Modeling for Quadruped Robot}
 \begin{figure}[th]
    \centering
    \includegraphics[width=0.4\textwidth]{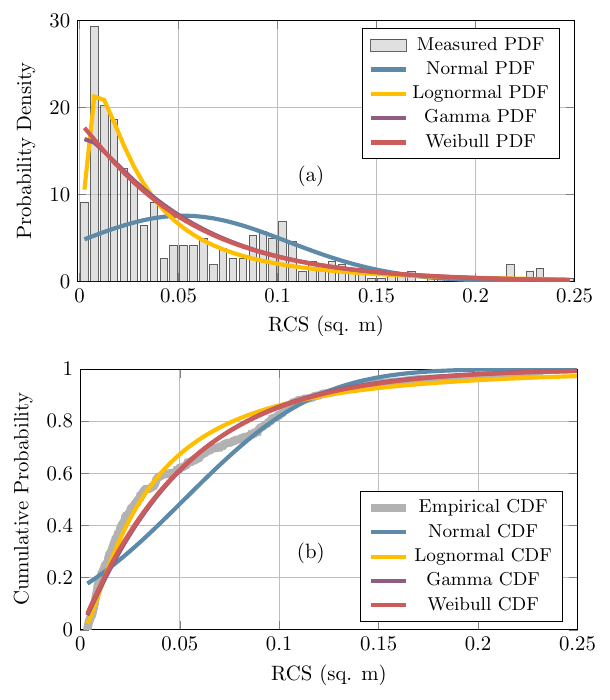} 
    \caption{RCS Distribution Modeling for QR at \(28\)GHz for \(\theta_b = 40^\circ\): (a) PDF Fitting, (b) CDF Fitting.}
    \label{robot_28_40_results} 
\end{figure}
 \begin{figure}[th]
    \centering
    \includegraphics[width=0.45\textwidth]{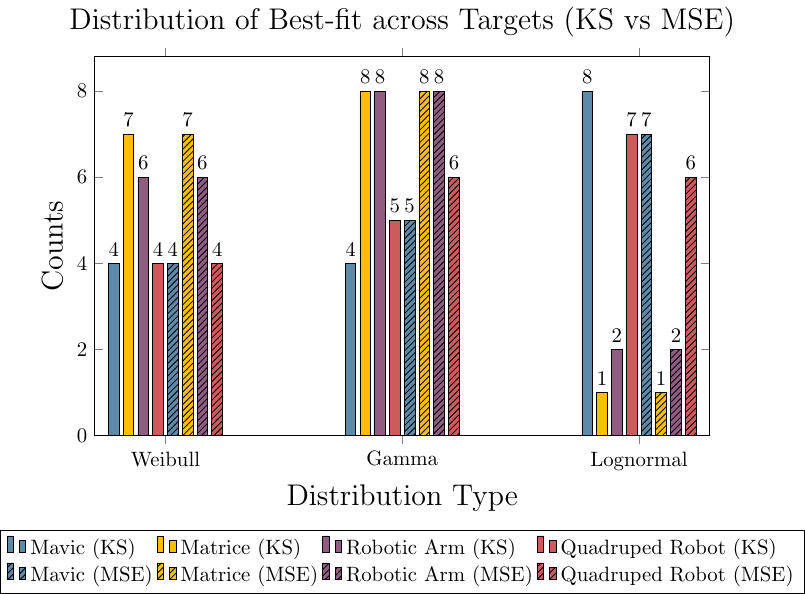} 
    \caption{\textcolor{black}{Unified win counts selected distributions over $25$-$28$ GHz and all bistatic angles.}}
    \label{summary_statistical} 
\end{figure}
 \begin{figure}[th]
    \centering
    \includegraphics[trim={5mm 0mm 3mm 3mm},clip,width=0.48\textwidth]{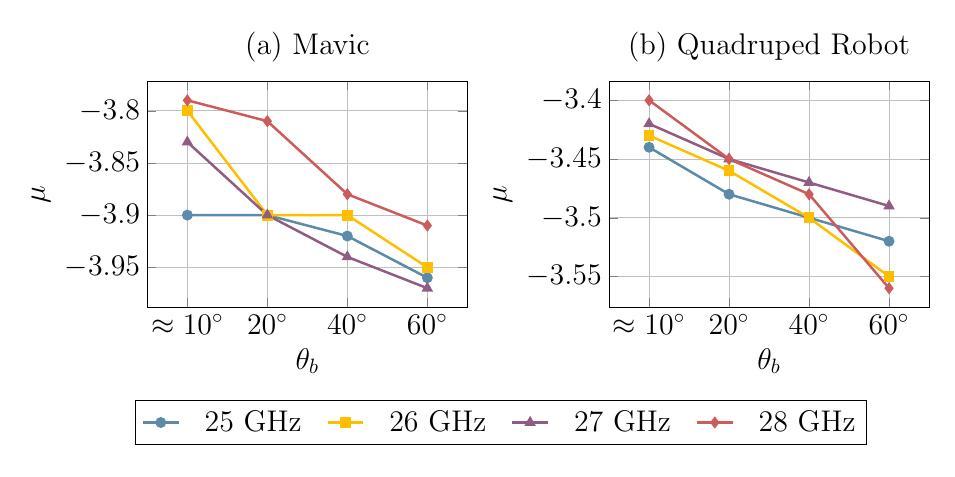} 
    \caption{\textcolor{black}{Impact of \(\theta_b\) and frequency on the \(\mu\) obtained using lognormal distribution for (a) Mavic \(2\) Pro and (b) \ac{QR}.}}
    \label{rcs_thetab_fc} 
\end{figure}
The fitting of the measured \ac{RCS} for \ac{QR} at \SI{28}{\GHz} and \(\theta_b = 40^\circ\) is depicted in Fig. \ref{robot_28_40_results}. The measured \ac{RCS} exhibits a positively skewed \ac{PDF}, characterized by a predominance of weak reflections and a limited number of strong backscatter reflections from the target, which is anticipated because \textcolor{black}{pose states} of the \ac{QR} introduce orientation-dependent variations in the \ac{RCS}. Consequently, strong reflections are observed when the robot's orientation favors specular backscatter, while weak reflections occur when the orientation reduces effective backscatter. The strong and weak backscatter also results in the bimodality observed in the \ac{PDF} of measured \ac{RCS} data.

Fig. \ref{robot_28_40_results}(a) shows the \ac{PDF} fitting of different parametric distributions to the measured \ac{RCS} \textcolor{black}{for the \ac{QR}}. The fitting results corroborated with the \ac{GoF} metrics are summarized in Tables \ref{robot_25}-\ref{robot_28} covering all \((\theta_b,f)\) combinations. The \ac{GoF}
metrics evaluating each distribution's accuracy are also provided the tables. The empirical \ac{CDF} fitting is provided in Fig. \ref{robot_28_40_results}(b) which also validates these results, demonstrating strong agreement between the empirical data and the fitted lognormal and Weibull distributions.  The following conclusions can be drawn from the results:
\begin{enumerate}
    \item \textcolor{black}{Based on \ac{KS} \ac{GoF}, the lognormal distribution provides the best fit to the measured \ac{RCS} in most representative cases. Specifically, it performs best in $7$ out of $16$ cases across different $(\theta_b,f)$ combinations. The Weibull and gamma distributions yield the best fit in $4$ and $5$ cases, respectively. In most cases, the differences in the \ac{GoF} metrics among the lognormal, Weibull, and gamma distributions are small.}
    \item It is observed that the \ac{RCS} decreases with an increase in \(\theta_b\) across all considered frequencies. Furthermore, the \ac{RCS} slightly increases with an increase in the operating frequency for a given \(\theta_b\). \textcolor{black}{These trends are evident in Fig. \ref{rcs_thetab_fc}(b).}
\end{enumerate}

\begin{table}[h!]
\centering
\small
\caption{Parameters at \SI{25}{\GHz} for varying \(\theta_b\) for QR.}
\begin{tabular}{|p{1.38cm}|p{1.05cm}|p{1.1cm}|p{3cm}|}
\hline
\multicolumn{4}{|c|}{\textbf{25 GHz, \(\theta_b \approx 10^\circ\)}} \\ \hline
\textbf{Dist.} & \textbf{KS Stat}\newline$(\times 10^{-2})$ & \textbf{MSE}\newline$(\times 10^{-3})$ & \textbf{Parameters} \\ \hline
Normal & \(18\) & \(10\) & \(\mu = 0.072, \sigma = 0.077\) \\
Lognormal & \(9.9\) & \(2.1\) & \(\mu = -3.44, \sigma = 1.52\) \\
\rowcolor{gray!20}
Gamma & \(5.1\) & \(0.8\) & \(A = 0.73, B = 0.099\) \\
Weibull & \(5.7\) & \(0.87\) & \(A = 0.065, B = 0.81\) 
\\ \hline

\multicolumn{4}{|c|}{\textbf{25 GHz, \(\theta_b =20^\circ\)}} \\ \hline
Normal & \(17\) & \(8.6\) & \(\mu = 0.055, \sigma = 0.054\) \\
Lognormal & \(11\) & \(2.6\) & \(\mu = -3.48, \sigma = 1.17\) \\
Gamma & \(9.3\) & \(2.7\) & \(A = 0.98, B = 0.056\) \\
\rowcolor{gray!20}
Weibull & \(8.9\) & \(2.5\) & \(A = 0.054, B = 0.98\) 
\\ \hline

\multicolumn{4}{|c|}{\textbf{25 GHz, \(\theta_b =40^\circ\)}} \\ \hline
Normal & \(17\) & \(11\) & \(\mu = 0.053, \sigma = 0.054\) \\
\rowcolor{gray!20}
Lognormal & \(5.2\) & \(0.86\) & \(\mu = -3.50, \sigma = 1.18\) \\
Gamma & \(5.9\) & \(0.86\) & \(A = 0.99, B = 0.054\) \\
\rowcolor{black!20}
Weibull & \(5.8\) & \(0.69\) & \(A = 0.053, B = 0.98\) 
\\ \hline

\multicolumn{4}{|c|}{\textbf{25 GHz, \(\theta_b =60^\circ\)}} \\ \hline
Normal & \(19\) & \(13\) & \(\mu = 0.046, \sigma = 0.048\) \\
\rowcolor{gray!20}
Lognormal & \(4.4\) & \(0.5\) & \(\mu = -3.52, \sigma = 0.98\) \\
Gamma & \(7.1\) & \(1.4\) & \(A = 1.22, B = 0.038\) \\
Weibull & \(7.3\) & \(1.3\) & \(A = 0.048, B = 1.07\) 
\\ \hline
\end{tabular}
\label{robot_25}
\end{table}
\begin{table}[h!]
\centering
\small
\caption{Parameters at \SI{26}{\GHz} for varying \(\theta_b\) for QR.}
\begin{tabular}{|p{1.38cm}|p{1.05cm}|p{1.1cm}|p{3cm}|}
\hline
\multicolumn{4}{|c|}{\textbf{26 GHz, \(\theta_b \approx 10^\circ\)}} \\ \hline
\textbf{Dist.} & \textbf{KS Stat}\newline$(\times 10^{-2})$ & \textbf{MSE}\newline$(\times 10^{-3})$ & \textbf{Parameters} \\ \hline
Normal & \(18\) & \(10\) & \(\mu = 0.055, \sigma = 0.056\) \\
\rowcolor{gray!20}
Lognormal & \(9.9\) & \(2\) & \(\mu = -3.44, \sigma = 1.10\) \\
Gamma & \(11\) & \(2.1\) & \(A = 1.04, B = 0.053\) \\
\rowcolor{black!20}
Weibull & \(10\) & \(1.7\) & \(A = 0.055, B = 0.99\) 
\\ \hline

\multicolumn{4}{|c|}{\textbf{26 GHz, \(\theta_b =20^\circ\)}} \\ \hline
Normal & \(15\) & \(6.2\) & \(\mu = 0.058, \sigma = 0.055\) \\
Lognormal & \(10\) & \(3\) & \(\mu = -3.46, \sigma = 1.3\) \\
\rowcolor{gray!20}
Gamma & \(7.9\) & \(1.3\) & \(A = 0.92, B = 0.063\) \\
Weibull & \(8.2\) & \(1.5\) & \(A = 0.057, B = 0.95\) 
\\ \hline

\multicolumn{4}{|c|}{\textbf{26 GHz, \(\theta_b =40^\circ\)}} \\ \hline
Normal & \(14\) & \(5.3\) & \(\mu = 0.058, \sigma = 0.055\) \\
Lognormal & \(11\) & \(3.8\) & \(\mu = -3.5, \sigma = 1.4\) \\
\rowcolor{gray!20}
Gamma & \(6.6\) & \(0.97\) & \(A = 0.87, B = 0.066\) \\
Weibull & \(6.7\) & \(1.1\) & \(A = 0.057, B = 0.94\) 
\\ \hline

\multicolumn{4}{|c|}{\textbf{26 GHz, \(\theta_b =60^\circ\)}} \\ \hline
Normal & \(20\) & \(14\) & \(\mu = 0.052, \sigma = 0.056\) \\
\rowcolor{gray!20}
Lognormal & \(5.7\) & \(1.1\) & \(\mu = -3.55, \sigma = 1.14\) \\
Gamma & \(10\) & \(3.2\) & \(A = 0.96, B = 0.054\) \\
Weibull & \(8.6\) & \(2.4\) & \(A = 0.05, B = 0.94\) 
\\ \hline
\end{tabular}
\label{robot_26}
\end{table}
\begin{table}[h!]
\centering
\small
\caption{Parameters at \SI{27}{\GHz} for varying \(\theta_b\) for QR.}
\begin{tabular}{|p{1.38cm}|p{1.05cm}|p{1.1cm}|p{3cm}|}
\hline
\multicolumn{4}{|c|}{\textbf{27 GHz, \(\theta_b \approx 10^\circ\)}} \\ \hline
\textbf{Dist.} & \textbf{KS Stat}\newline$(\times 10^{-2})$ & \textbf{MSE}\newline$(\times 10^{-3})$ & \textbf{Parameters} \\ \hline
Normal & \(16\) & \(7.1\) & \(\mu = 0.062, \sigma = 0.062\) \\
Lognormal & \(14\) & \(4.2\) & \(\mu = -3.42, \sigma = 1.3\) \\
\rowcolor{black!20}
Gamma & \(9.2\) & \(2.2\) & \(A = 0.89, B = 0.069\) \\
\rowcolor{gray!20}
Weibull & \(9.1\) & \(2.3\) & \(A = 0.06, B = 0.93\) 
\\ \hline

\multicolumn{4}{|c|}{\textbf{27 GHz, \(\theta_b =20^\circ\)}} \\ \hline
Normal & \(15\) & \(8.5\) & \(\mu = 0.069, \sigma = 0.068\) \\
Lognormal & \(11\) & \(2.9\) & \(\mu = -3.45, \sigma = 1.53\) \\
\rowcolor{gray!20}
Gamma & \(7.9\) & \(1.3\) & \(A = 0.75, B = 0.092\) \\
Weibull & \(8.4\) & \(1.6\) & \(A = 0.064, B = 0.84\) 
\\ \hline

\multicolumn{4}{|c|}{\textbf{27 GHz, \(\theta_b =40^\circ\)}} \\ \hline
Normal & \(14\) & \(5\) & \(\mu = 0.056, \sigma = 0.052\) \\
Lognormal & \(10\) & \(3.6\) & \(\mu = -3.47, \sigma = 1.29\) \\
\rowcolor{gray!20}
Gamma & \(6.9\) & \(1.6\) & \(A = 0.95, B = 0.059\) \\
Weibull & \(7.1\) & \(1.7\) & \(A = 0.056, B = 0.98\) 
\\ \hline

\multicolumn{4}{|c|}{\textbf{27 GHz, \(\theta_b =60^\circ\)}} \\ \hline
Normal & \(22\) & \(17\) & \(\mu = 0.059, \sigma = 0.067\) \\
\rowcolor{gray!20}
Lognormal & \(10\) & \(3.8\) & \(\mu = -3.49, \sigma = 1.16\) \\
Gamma & \(15\) & \(7\) & \(A = 0.88, B = 0.067\) \\
Weibull & \(13\) & \(5.5\) & \(A = 0.055, B = 0.88\) 
\\ \hline
\end{tabular}
\label{robot_27}
\end{table}
\begin{table}[h!]
\centering
\small
\caption{Parameters at \SI{28}{\GHz} for varying \(\theta_b\) for QR.}
\begin{tabular}{|p{1.38cm}|p{1.05cm}|p{1.1cm}|p{3.2cm}|}
\hline
\multicolumn{4}{|c|}{\textbf{28 GHz, \(\theta_b \approx 10^\circ\)}} \\ \hline
\textbf{Dist.} & \textbf{KS Stat}\newline$(\times 10^{-2})$ & \textbf{MSE}\newline$(\times 10^{-3})$ & \textbf{Parameters} \\ \hline
Normal & \(15\) & \(7\) & \(\mu = 0.06, \sigma = 0.057\) \\
Lognormal & \(8.8\) & \(2\) & \(\mu = -3.4, \sigma = 1.22\) \\
Gamma & \(6.7\) & \(1.2\) & \(A = 0.97, B = 0.061\) \\
\rowcolor{gray!20}
Weibull & \(6.6\) & \(1.2\) & \(A = 0.059, B = 0.98\) 
\\ \hline

\multicolumn{4}{|c|}{\textbf{28 GHz, \(\theta_b =20^\circ\)}} \\ \hline
Normal & \(18\) & \(9.9\) & \(\mu = 0.056, \sigma = 0.053\) \\
\rowcolor{black!20}
Lognormal & \(10\) & \(1.8\) & \(\mu = -3.45, \sigma = 1.19\) \\
Gamma & \(9.9\) & \(2.3\) & \(A = 0.99, B = 0.056\) \\
\rowcolor{gray!20}
Weibull & \(9.9\) & \(2.2\) & \(A = 0.056, B = 0.99\) 
\\ \hline

\multicolumn{4}{|c|}{\textbf{28 GHz, \(\theta_b =40^\circ\)}} \\ \hline
Normal & \(18\) & \(10\) & \(\mu = 0.0523, \sigma = 0.052\) \\
\rowcolor{gray!20}
Lognormal & \(8.1\) & \(1.5\) & \(\mu = -3.48, \sigma = 1.09\) \\
Gamma & \(9.5\) & \(2.5\) & \(A = 1.07, B = 0.048\) \\
Weibull & \(8.6\) & \(2.1\) & \(A = 0.052, B = 1.01\) 
\\ \hline

\multicolumn{4}{|c|}{\textbf{28 GHz, \(\theta_b =60^\circ\)}} \\ \hline
Normal & \(21\) & \(15\) & \(\mu = 0.05, \sigma = 0.05\) \\
\rowcolor{gray!20}
Lognormal & \(5.7\) & \(0.9\) & \(\mu = -3.56, \sigma = 1.09\) \\
Gamma & \(9.8\) & \(3.4\) & \(A = 1.0038, B = 0.05\) \\
Weibull & \(8.6\) & \(2.5\) & \(A = 0.049, B = 0.96\) 
\\ \hline
\end{tabular}
\label{robot_28}
\end{table}
%
%
\subsection{Unified RCS Distribution}\label{sec3d}
\textcolor{black}{In addition to Tables \ref{mavic_25}-\ref{robot_28}, we report a cross-scenario aggregation of the best-fitting distribution families across all \((\theta_b,f)\) combinations.
Our analysis presented in Fig. \ref{summary_statistical} uses win count statistics, where for each target \(t\), we consider the \(16\) scenarios given by all frequency-angle, i.e. \((f,\theta_b)\), combinations.  In scenario \(i\), we fit the four candidate families, i.e., normal, \(\mathcal{N}\), lognormal, \(\mathcal{LN}\), Weibull, \(\mathcal{W}\), and gamma, \(\Gamma\) and select the best-fitting one according to either the \(\mathrm{KS}\) statistic or the \(\mathrm{MSE}\) criterion (denoted \(c\in\{\mathrm{KS},\mathrm{MSE}\}\)) denoted by \(d_{i}^{(t,c)} \in \{\mathcal{N},\mathcal{LN},\mathcal{W},\Gamma\}\), for \(i=1,\dots,16\). 
To synthesize the results from Tables \ref{mavic_25}-\ref{robot_28}, we aggregate the number of times each family $d \in\{\mathcal{N}, \mathcal{L} \mathcal{N}, \mathcal{W}, \Gamma\}$ provides the best fit across the $16$ parameter scenarios. We calculate the win count as $N_d^{(t, c)}= \sum_{i=1}^{16} \mathbf{1}\left[d_i^{(t, c)}=d\right]$ and visualize the results in Fig. \ref{summary_statistical}. The histograms display the performance for each of the four targets, separated by metric (KS and MSE).
Fig. \ref{summary_statistical} indicates that within the considered measurement scope and validity domain, different targets are best described by different statistical distributions, which is physically meaningful and can be explained by differences in target size, structural complexity, material composition, and motion dynamics. 
Smaller and more agile targets, such as the Mavic \(2\) Pro and the \ac{QR}, experience rapid orientation changes and exhibit reflections spanning several orders of magnitude due to intermittent dominance of a few scattering centers. The rapid movements lead to heavier right-tailed distributions, making lognormal/Weibull statistics more suitable. In contrast, larger and more structurally complex platforms, such as the Matrice \(300\) RTK and the \ac{RA}, contain a greater number of contributing scatterers across aspect/pose variations. The increased scattering diversity tends to average extreme returns, resulting in \ac{RCS} statistics that are more consistently captured by the gamma distribution.}

Interestingly, both distributions are special instances of the more general \textit{generalized Gamma distribution}, whose \ac{PDF} is:
\begin{equation}
	f(x ; a, d, p)=\frac{p}{a^d \Gamma(d / p)} x^{d-1} e^{-(x / a)^p}, \quad x>0,
\end{equation} 
where $a,d,p$ denote the scale, shape, and power parameters and $\Gamma(\cdot)$ is the gamma function.
When $p = 1$, we get the gamma distribution and when $p \rightarrow 0$, we get the lognormal distribution.
 $p$ controls the skewness of the distribution. 
\subsection{\textcolor{black}{Statistical \ac{RCS} Integration to Target Channel Path Loss}}\label{sec3e}
\textcolor{black}{This section elaborates on how the obtained \ac{RCS} statistics can be incorporated in the target channel \ac{PL}. Although Section \ref{sec3} focuses on modeling the stochastic scattering term $\sigma$, the fitted distributions can be directly mapped into a target channel \ac{PL} term via the bistatic radar equation. From (\ref{radar_equation}), the target-induced received power is $P_{\mathrm{tar}} = C(f,\theta_b,d)\,\sigma$,
where $C(f,\theta_b,d)\triangleq \frac{P_t G_{\mathrm{Tx}}G_{\mathrm{Rx}}\,\lambda^2 L}{(4\pi)^3 d^4}$ collects geometry- and system-dependent factors. Equivalently, considering \(d_\mathrm{Tx,tar} = d_\mathrm{tar,Rx} = d\), the \ac{ISAC} target channel loss can be defined as:
\begin{equation}\label{target_chan_loss}
L_{\mathrm{tar}}(f,\theta_b,d)\triangleq \frac{(4\pi)^3 d^4}{G_{\mathrm{Tx}}G_{\mathrm{Rx}}\lambda^2 L\,\sigma},
\end{equation}
which yields:
\begin{equation}\label{target_chan_loss2}
L_{\mathrm{tar,dB}} = 10\log_{10}\!\Big(\frac{(4\pi)^3 d^4}{G_{\mathrm{Tx}}G_{\mathrm{Rx}}\lambda^2 L}\Big) - 10\log_{10}(\sigma).
\end{equation}
Therefore, for fixed $(f,\theta_b,d)$ the object-type/frequency/$\theta_b$ dependence of the target link budget is governed by the distribution of $\sigma$ (Tables \ref{mavic_25}-\ref{robot_28}), while the  propagation environment can be modeled using established channel models.}

\textcolor{black}{This section provided a comprehensive statistical \ac{RCS} characterization of \ac{InF} targets that is directly suitable for standardization and Monte-Carlo \ac{ISAC} channel simulations, where the target contribution can be represented through a random variable. However, this abstraction does not explicitly capture range and \(\theta_b\) dependent structure that becomes critical when either the geometry is known, e.g., digital-twin/site-specific sensing or when operating in the \ac{NF} regime. Motivated by this, we next introduce a deterministic framework that estimates \ac{RCS} by coupling measured two-way \ac{PL} with \ac{RCS} models for a canonical reflector.}
\subsection{Application of \ac{RCS} Measurements} Beyond \ac{RCS} modeling parameterization, the reported best-fit distributions can also serve as reference signatures for online \textit{target classification}. In a deployed \ac{ISAC} receiver, instantaneous \ac{RCS} samples can be accumulated over a finite time window, and an empirical distribution can be constructed from the resulting ensemble. The empirical distribution is then compared against the catalogue of target-class distributions reported herein using a divergence metric such as the Kullback-Leibler (KL) divergence \cite{cheng2019classification}, with the candidate target-class distribution yielding the minimum divergence selected as the inferred target class. The described KL-based target classification procedure provides a statistical classification mechanism that exploits the target-dependent shape and scale parameters identified in this study. In addition, the KL-based target classification procedure complements deterministic angle-based sensing by enabling target-type inference.
\section{Framework for Deterministic Radar Cross Section Evaluation}\label{sec4}
While Section \ref{sec3} provides target-class \acp{PDF} suitable for system-level modeling, deterministic modeling is required when the target-dependent channel term must be conditioned on the \ac{Tx}-target-\ac{Rx} geometry or when \ac{NF} effects invalidate far-field assumptions. In these regimes, the goal is to determine explicit \ac{RCS} characterization that captures structured angular/range dependence. The objective of the deterministic framework is to obtain a configuration-aware, physically interpretable \ac{RCS} model that captures systematic \ac{NF} bistatic dependencies under the measurement domain and validity defined for deterministic \ac{RCS} characterization in Section \ref{sec2c}. Accordingly, this section outlines a framework for the deterministic evaluation of the bistatic \ac{RCS} in the \ac{NF}. Specifically, the framework is designed to characterize the \ac{RCS} of a rectangular sheet of laminated wood in \ac{NF} bistatic configuration, operating within a frequency range of \SI{25}{}-\SI{28}{\GHz}. The methodology integrates empirical \ac{PL} data with different \ac{RCS} models, providing a \textcolor{black}{\ac{FI} and \ac{CI}-based} fitting process to derive the target's \ac{RCS}. The Rayleigh distance for the target is calculated as \(D = \frac{2S^2}{\lambda}\) \cite{4154040}, where \(D\) represents the Rayleigh distance, \(S\) is the largest dimension of the target (\SI{1.84}{\meter} in this case), and \(\lambda\) is the operating wavelength. For the given frequency range of \SI{25}{}-\SI{28}{\GHz}, the selected target's Rayleigh distance lies approximately between \SI{564}{\meter} and \SI{632}{\meter}. 
\vspace{-5mm}
\subsection{Measurement Setup}\label{sec4a}
\textcolor{black}{In the \ac{InH} measurement setup, the midpoint of the line segment joining the \ac{Tx} and \ac{Rx} is located at $(0,0)$. The \ac{Tx} and \ac{Rx} are placed at $(-a,0)$ and $(a,0)$, respectively, with $a=0.7$~m. The target is positioned at $(0,y)$, where $y\in[2,10]$~m. Owing to symmetry, the \ac{Tx}-target and \ac{Rx}-target distances are equal and given by \(
d \triangleq d_{\mathrm{Tx,tar}} = d_{\mathrm{Rx,tar}} = \sqrt{a^2+y^2}\). Considering \ac{Tx}-target and \ac{Rx}-target geometry and using the law of cosines, the cosine of \(\theta_b\) is given as:
\begin{equation}
\cos\theta_b = 1 - 2\left(\frac{a}{d}\right)^2,
\end{equation}
and therefore, \(\theta_b\) in degrees can be expressed as:
\begin{equation}
\theta_b(d)=\arccos\!\left(1-2\left(\frac{a}{d}\right)^2\right)\cdot\frac{180^\circ}{\pi}. \tag{5}
\end{equation}
Equivalently, since $d=\sqrt{a^2+y^2}$, one may also write as:
\begin{equation}
\theta_b(y)=\arccos\!\left(\frac{y^2-a^2}{y^2+a^2}\right)\cdot\frac{180^\circ}{\pi}.
\end{equation}
For $y\in[2,10]$~m (i.e., $d\in[\sqrt{a^2+2^2},\sqrt{a^2+10^2}]$), $\theta_b$ decreases from approximately $38.6^\circ$ to $8.0^\circ$.}

\textcolor{black}{In the deterministic \ac{RCS} measurements, the rectangular laminated wood sheet is oriented such that the surface normal bisects the \ac{Tx}-target and target-\ac{Rx} directions, so that the \ac{Rx} receives the dominant specular bistatic return for each target position varied through \(d\). 
The specific orientation and setup of the wood sheet is a deliberate experimental design choice because the \ac{RCS} identification is best-posed when the target-induced component dominates the received power with sufficiently high \ac{SCNR}, thereby minimizing bias from uncontrolled environment-dependent multipath that would otherwise be absorbed into the fitted \ac{RCS} term.} \textcolor{black}{If \ac{Rx} moves away from the specular direction of the echo, the received signal may be dominated by diffuse/edge scattering and multipath, which reduces \ac{SCNR} and complicates the identification of \ac{RCS} from \ac{PL} measurements.}
 \begin{figure}[t]
    \centering
    \includegraphics[width=0.36\textwidth]{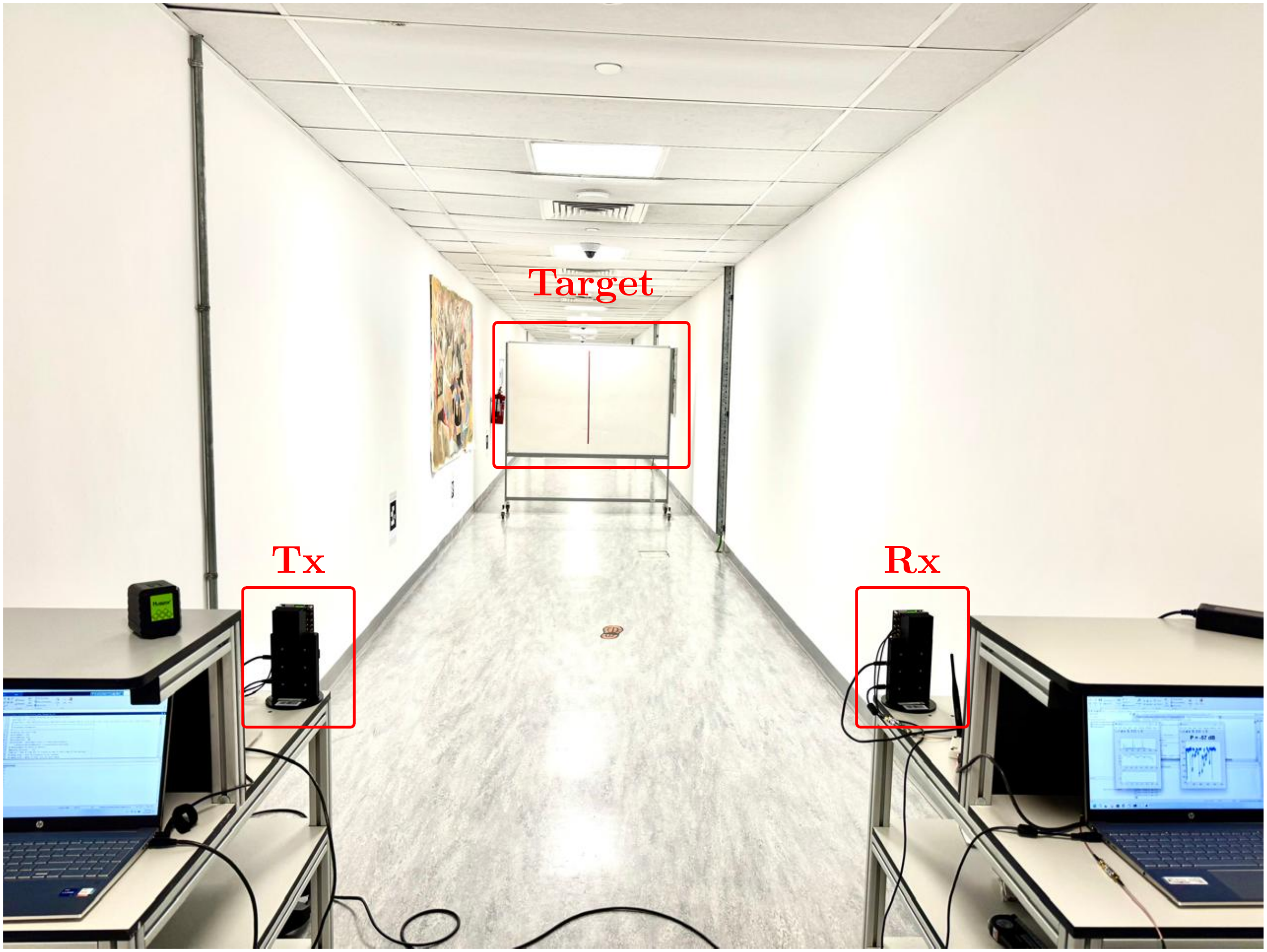} 
    \caption{The {\ac{InH}} scenario for deterministic RCS evaluation.}
    \label{inh_setup} 
\end{figure}
\vspace{-5mm}
\subsection{RCS Analysis}\label{sec4b}
\textcolor{black}{The modified \ac{FI} and \ac{CI} model based \ac{PL} equations for bistatic system are given as:}
\begin{equation}\label{pl_model_fi}
   \textcolor{black}{ \mathrm{PL}^\mathrm{FI}_i(d,\lambda) = \alpha + 20n\log_{10}(d) -10\log_{10}(\sigma_i(d,\lambda)) + X_\sigma^\mathrm{FI},}
\end{equation}
and 
\begin{equation}\label{pl_model_ci}
\begin{split}
       \textcolor{black}{ \mathrm{PL}^\mathrm{CI}_i(d,\lambda)} =&\textcolor{black}{ \mathrm{FSPL}(d_0,\lambda) + 20n\log_{10}\left(\frac{d}{d_0}\right)}\\
       &\textcolor{black}{-10\log_{10}(\sigma_i(d,\lambda)) + X_\sigma^\mathrm{CI},}
       \end{split}
\end{equation}
\textcolor{black}{respectively. In \eqref{pl_model_ci}, \(\alpha\) is the intercept acting as a bias, \(n\) is the \ac{PL} exponent, \(\sigma_i(d,\lambda)\) is the \(i\)th \textcolor{black}{\(\theta_b\)-dependent} \ac{RCS} model, and \(X^\mathrm{FI}_\sigma\) captures the shadowing effect in case of \ac{FI} model. On the other hand, in \eqref{pl_model_ci}, \(d_0\) is a physically-based reference distance, and
\(\mathrm{FSPL}(d_0,\lambda)\) is the reference free-space path loss value at \(d_0=\) \SI{1}{\meter}. Moreover, \(X^\mathrm{CI}_\sigma\) captures the shadowing effect in case of \ac{CI} model} Similar to \cite{108383}, \textcolor{black}{for both models,} the term in the single-way \ac{PL}, i.e., $10n\log_{10}(d)$ is accounted for twice due to the two-way propagation, i.e., \ac{Tx} towards target, then target towards \ac{Rx}. 

Following \cite{4154040}, the \ac{NF} \ac{RCS} of the rectangular sheet depends on both \(d\) and \(\lambda\), because in the \ac{NF}, the of energy reflected back towards the \ac{Rx} varies with \(d\), as the signal may not fully illuminate the target at different \ac{NF} distances. Accordingly, we explore multiple \ac{NF} \ac{RCS} models for the rectangular sheet as:
\begin{equation}\label{approx_rcs}
\begin{split}
\sigma_1(d,\lambda) &= a_1d^2\cos^m(\theta_b(d)),\\
\sigma_2(d,\lambda) &= \left(a_1d^2+a_2\lambda d^3\right)\cos^m(\theta_b(d)),\\
\sigma_3(d,\lambda) &= \left(a_1d^2+a_2\lambda d^3+a_3\lambda^2 d^4\right)\cos^m(\theta_b(d)),
\end{split}
\end{equation}
where \( \sigma_1(d, \lambda) \), \( \sigma_2(d, \lambda) \), and \( \sigma_3(d, \lambda) \) are the proposed bistatic \ac{RCS} models for the rectangular sheet as functions of \(d\) and \(\lambda\). \( a_1, a_2, \) and \( a_3 \) are fitting parameters that characterize the \ac{RCS} based on the geometric properties of the target. Note that (\ref{approx_rcs}) integrates the \(\theta_b\) dependence into the monostatic \ac{RCS} models derived from \cite{4154040}. However, the models in (\ref{approx_rcs}) can be used only for \(\theta_b \in \left(0^\circ,90^\circ\right)\). 

As emphasized in Section \ref{sec2c}, we have used only one rectangular sheet for our measurement campaign, we cannot characterize the dependence of \( a_1, a_2, \) and \( a_3 \) on the area of the sheet. \( \cos^m(\theta_b(d)) \) captures the angular dependence of the bistatic \ac{RCS}, where \(m\) is an angular exponent. \( d^2, \lambda d^3, \) and \( \lambda^2 d^4 \) represent the terms that account for the \ac{NF} scattering behavior of the rectangular sheet at different \(d\) and \(\lambda\).

To evaluate the \ac{GoF}, \ac{MFE} is evaluated for each \ac{RCS} model, where \ac{MFE} in percentage is defined as:
\begin{equation}
  \textcolor{black}{\mathrm{MFE} = \frac{1}{N} \sum\nolimits_{j=1}^N \left| \frac{\mathrm{PL}_{\mathrm{meas},j} - \mathrm{PL}_{i,j}^{(\cdot)}}{\mathrm{PL}_{\mathrm{meas},j}} \right| \times 100\%}
\end{equation}
where \(N\) is the total number of data points, \(\mathrm{PL}_{\mathrm{meas},j}\) is the measured \ac{PL} for the \(j\)-th data point, \textcolor{black}{\(\mathrm{PL}_{i,j}^{(\cdot)}\) is the \(i\)th modeled \ac{PL} through either \ac{FI} or \ac{CI} model} for the \(j\)-th data point. \textcolor{black}{\((\cdot)\) in the superscript is used to differentiate between the \ac{FI} and \ac{CI} models.} The distinction among the various \ac{PL} models \textcolor{black}{(both in case of \ac{FI} and \ac{CI} models)} lies in the selection of the \ac{RCS} model \textcolor{black}{as given in \eqref{approx_rcs}. \(X_\sigma^{(\cdot)}\) for a given model} is evaluated as \(\sqrt{\frac{1}{N} \sum\nolimits_{j=1}^N (x_j - \mu)^2}\), where \(x_j\) represents the difference between the measured values and the fitted curve values \textcolor{black}{for the respective model}, and \(\mu\) denotes the average value of \(x_j\). 

\begin{figure}[t]
    \centering
    \includegraphics[trim={3mm 5mm 2mm 0mm},clip,width=0.5\textwidth]{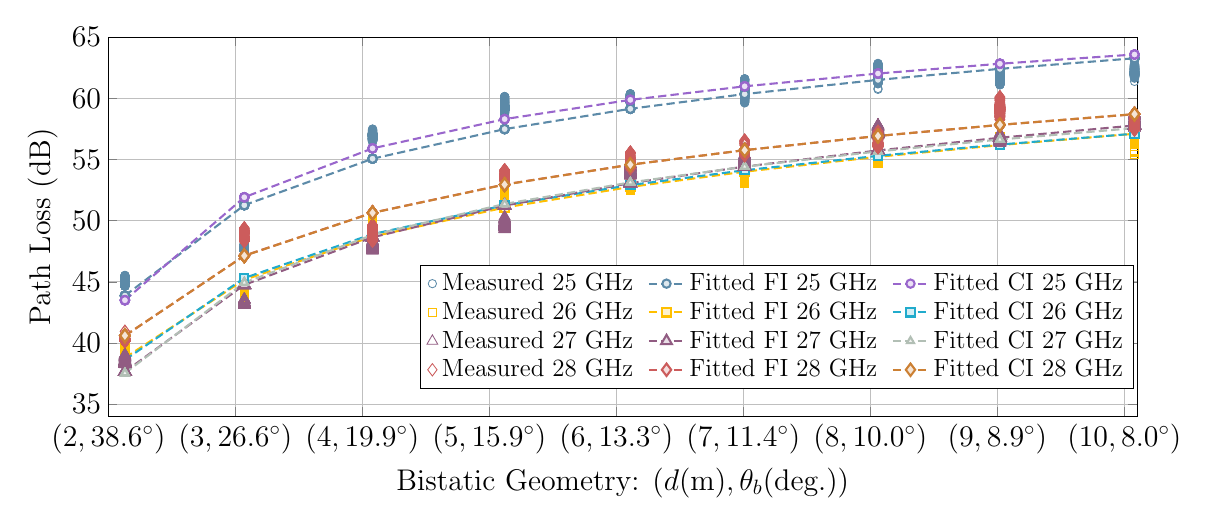} 
    \caption{Measured \ac{PL} and model curve fitting \textcolor{black}{using \ac{FI} and \ac{CI} models} in the \ac{NF}. \textcolor{black}{\ac{Tx}/\ac{Rx} are arranged to observe the dominant specular lobe of the rectangular sheet.}} 
    \label{result_deter_rcs} 
\end{figure}

\textcolor{black}{To justify \eqref{pl_model_fi} and \eqref{pl_model_ci}}, the inverse of \ac{PL}, i.e., path gain as a function of \textcolor{black}{bistatic geometry, i.e., \((d,\theta_b)\)}, is plotted in Fig. \ref{result_deter_rcs}, from which we observe that the \textcolor{black}{measured path gain curves do not} follow a linear growth pattern but instead reach a plateau as \(d\) increases. The plateau is a characteristic of the \ac{NF} regime, arising from the non-linear behavior of the \ac{NF} \ac{RCS} of the target. The trend remains consistent across \SI{25}{}-\SI{28}{\GHz} \textcolor{black}{band employed in the measurement campaign}. 

\begin{table*}[t]
\centering
\caption{Fitting parameters for deterministic RCS using the FI and CI models.}
\footnotesize
\setlength{\tabcolsep}{3pt}
\renewcommand{\arraystretch}{1.1}
\resizebox{\textwidth}{!}{
\begin{tabular}{|c|c|c|c|c|c|c|c|c|c| c |c|c|c|c|c|c|c|c|c|c|}
\cline{1-10} \cline{12-21}

\multicolumn{10}{|c|}{\textbf{25 GHz}} & & \multicolumn{10}{|c|}{\textbf{26 GHz}} \\ \cline{1-10} \cline{12-21}
\textbf{Model} & \textbf{Approx.} & \(\alpha\) & \(n\) & \(m\) & \(a_1\) & \(a_2\) & \(a_3\) & \(X_\sigma^{(\cdot)}\) & \textbf{MFE} (\(\%\)) & & \textbf{Model} & \textbf{Approx.} & \(\alpha\) & \(n\) & \(m\) & \(a_1\) & \(a_2\) & \(a_3\) & \(X_\sigma^{(\cdot)}\) & \textbf{MFE} (\(\%\)) \\ \cline{1-10} \cline{12-21}

FI & \(\sigma_1(d,\lambda)\) & \(51.41\) & \(1.85\) & \(-7.86\) & \(2.96\) & \(-\)   & \(-\)   & \(1.64\) & \(2.40\) & & FI & \(\sigma_1(d,\lambda)\) & \(44.21\) & \(1.90\) & \(-6.10\) & \(2.94\) & \(-\)   & \(-\)   & \(0.72\) & \(1.15\) \\ \cline{1-10} \cline{12-21}
FI & \(\sigma_2(d,\lambda)\) & \(51.66\) & \(1.84\) & \(-8.13\) & \(2.91\) & \(1.25\) & \(-\)   & \(1.63\) & \(2.38\) & & FI & \(\sigma_2(d,\lambda)\) & \(43.80\) & \(1.92\) & \(-6.02\) & \(2.90\) & \(1.25\) & \(-\)   & \(0.72\) & \(1.14\) \\ \cline{1-10} \cline{12-21}
FI & \(\sigma_3(d,\lambda)\) & \(51.82\) & \(1.83\) & \(-8.26\) & \(2.90\) & \(1.25\) & \(0.03\) & \(1.63\) & \(2.37\) & & FI & \(\sigma_3(d,\lambda)\) & \(43.65\) & \(1.92\) & \(-6.02\) & \(2.80\) & \(1.25\) & \(0.03\) & \(0.72\) & \(1.14\) \\ \cline{1-10} \cline{12-21}
CI & \(\sigma_1(d,\lambda)\) & \(-\) & \(1.73\) & \(-10.39\) & \(4.51\) & \(-\) & \(-\) & \(1.67\) & \(2.48\) & & CI & \(\sigma_1(d,\lambda)\) & \(-\) & \(1.83\) & \(-6.96\) & \(4.33\) & \(-\) & \(-\) & \(0.73\) & \(1.17\) \\ \cline{1-10} \cline{12-21}
CI & \(\sigma_2(d,\lambda)\) & \(-\) & \(1.70\) & \(-10.88\) & \(4.31\) & \(1.73\) & \(-\) & \(1.67\) & \(2.44\) & & CI & \(\sigma_2(d,\lambda)\) & \(-\) & \(1.87\) & \(-6.27\) & \(4.22\) & \(1.01\) & \(-\) & \(0.73\) & \(1.17\) \\ \cline{1-10} \cline{12-21}
CI & \(\sigma_3(d,\lambda)\) & \(-\) & \(1.69\) & \(-10.79\) & \(4.43\) & \(1.83\) & \(0.04\) & \(1.66\) & \(2.40\) & & CI & \(\sigma_3(d,\lambda)\) & \(-\) & \(1.86\) & \(-6.86\) & \(4.45\) & \(1.25\) & \(0.05\) & \(0.73\) & \(1.19\) \\ \cline{1-10} \cline{12-21}

\multicolumn{21}{c}{} \\[-0.5em] \cline{1-10} \cline{12-21}

\multicolumn{10}{|c|}{\textbf{27 GHz}} & & \multicolumn{10}{|c|}{\textbf{28 GHz}} \\ \cline{1-10} \cline{12-21}
\textbf{Model} & \textbf{Approx.} & \(\alpha\) & \(n\) & \(m\) & \(a_1\) & \(a_2\) & \(a_3\) & \(X_\sigma^{(\cdot)}\) & \textbf{MFE} (\(\%\)) & & \textbf{Model} & \textbf{Approx.} & \(\alpha\) & \(n\) & \(m\) & \(a_1\) & \(a_2\) & \(a_3\) & \(X_\sigma^{(\cdot)}\) & \textbf{MFE} (\(\%\)) \\ \cline{1-10} \cline{12-21}

FI & \(\sigma_1(d,\lambda)\) & \(43.78\) & \(1.95\) & \(-7.19\) & \(2.99\) & \(-\)   & \(-\)   & \(2.09\) & \(1.81\) & & FI & \(\sigma_1(d,\lambda)\) & \(46.80\) & \(1.85\) & \(-6.56\) & \(3.02\) & \(-\)   & \(-\)   & \(1.04\) & \(1.60\) \\ \cline{1-10} \cline{12-21}
FI & \(\sigma_2(d,\lambda)\) & \(44.17\) & \(1.97\) & \(-7.04\) & \(3.50\) & \(1.01\) & \(-\)   & \(2.03\) & \(1.79\) & & FI & \(\sigma_2(d,\lambda)\) & \(47.73\) & \(1.85\) & \(-6.68\) & \(3.25\) & \(1.01\) & \(-\)   & \(1.04\) & \(1.58\) \\ \cline{1-10} \cline{12-21}
FI & \(\sigma_3(d,\lambda)\) & \(43.99\) & \(1.98\) & \(-6.91\) & \(3.50\) & \(1.01\) & \(0.02\) & \(2.02\) & \(1.77\) & & FI & \(\sigma_3(d,\lambda)\) & \(47.56\) & \(1.85\) & \(-6.69\) & \(3.50\) & \(1.01\) & \(0.02\) & \(1.04\) & \(1.58\) \\ \cline{1-10} \cline{12-21}
CI & \(\sigma_1(d,\lambda)\) & \(-\) & \(1.89\) & \(-8.16\) & \(4.21\) & \(-\) & \(-\) & \(2.14\) & \(1.95\) & & CI & \(\sigma_1(d,\lambda)\) & \(-\) & \(1.84\) & \(-6.74\) & \(3.61\) & \(-\)& \(-\) & \(1.04\) & \(1.59\) \\ \cline{1-10} \cline{12-21}
CI & \(\sigma_2(d,\lambda)\) & \(-\) & \(1.93\) & \(-8.35\) & \(4.55\) & \(2.01\) & \(-\) & \(2.09\) & \(2.03\) & & CI & \(\sigma_2(d,\lambda)\) & \(-\) & \(1.85\) & \(-6.68\) & \(3.12\) & \(1.01\) & \(-\)& \(1.04\) & \(1.59\) \\ \cline{1-10} \cline{12-21}
CI & \(\sigma_3(d,\lambda)\) & \(-\) & \(1.91\) & \(-7.83\) & \(3.88\) & \(1.91\) & \(0.018\) & \(2.07\) & \(1.96\) & & CI & \(\sigma_3(d,\lambda)\) & \(-\) & \(1.85\) & \(-6.64\) & \(4.08\) & \(1.01\) & \(0.018\) & \(1.04\) & \(1.60\) \\ \cline{1-10} \cline{12-21}

\end{tabular}
}
\label{rcs_dterm}
\end{table*}

Table \ref{rcs_dterm} presents the results of the fitting \textcolor{black}{parameters} for deterministic RCS models \(\sigma_1(d,\lambda), \sigma_2(d,\lambda)\), and \(\sigma_3(d,\lambda)\) across operating frequencies of \SI{25}{}-\SI{28}{\GHz} used in the measurement campaign \textcolor{black}{for both \ac{FI} and \ac{CI} models. For both models,} \(\sigma_3(d,\lambda)\) generally provides the lowest \ac{MFE}, due to its inclusion of higher-order terms. Moreover, the \ac{MFE} values decrease with increasing frequency, indicating improved fitting accuracy at higher frequencies. \textcolor{black}{For the \ac{FI} model,} the intercept, \(\alpha\) and \ac{PL} exponent, \(n\), show minor variations across frequencies and models, with \(\alpha\) decreasing slightly as frequency increases e.g., from \(51.41\) at \SI{25}{\GHz} to \(47.56\) at \SI{28}{\GHz}, while \(n\) remains consistent, ranging from \(1.83\) to \(1.98\). The angular exponent, \(m\), varies across frequencies, ranging from \(-8.26\) at \SI{25}{\GHz} for \(\sigma_3(d,\lambda)\) to \(-6.69\) at \SI{28}{\GHz} for \(\sigma_3(d,\lambda)\), indicating its sensitivity to frequency. \textcolor{black}{For the \ac{CI} model, \(n\) remains comparatively stable across frequency, taking values in the range \(1.73\)-\(1.89\). Similar to the \ac{FI} case, the angular parameter \(m\) exhibits a clear frequency dependence, varying from \(-10.79\) for \(\sigma_3(d,\lambda)\) at \SI{25}{GHz} to \(-6.64\) for \(\sigma_3(d,\lambda)\) at \SI{28}{GHz}.} The \ac{RCS} model parameters, \(a_1, a_2\), and \(a_3\) are consistent within each frequency and \textcolor{black}{both models}. The small values of the parameters \(a_2\) and \(a_3\) indicate their minor contributions towards the deterministic \ac{RCS}. However, it is again highlighted that \(a_2\) and \(a_3\) also depend on the area of the rectangular sheet. The change in the area of the sheet would certainly change \(a_2\) and \(a_3\). Specifically, \(a_2\) and \(a_3\) are expected to be higher for rectangular sheets with smaller areas. Note that while lower-order polynomials in (\ref{approx_rcs})  suffice for our test case, the optimal order may vary with target dimensions. The parameters \((m, a_1, a_2\) and \(a_3\)) are inherently dependent on the size of the rectangular sheet, meaning that larger or smaller targets may require higher-order terms to maintain fitting accuracy.
\textcolor{black}{\subsection{Impact of Change in \ac{RCS} on Path Loss}}\label{sec4c}
\textcolor{black}{From both \ac{PL} models, \eqref{pl_model_fi} and \eqref{pl_model_ci}, the target \ac{RCS} affects the measured \ac{PL} only through the $-10\log_{10}\bigl(\sigma_i(d,\lambda)\bigr)$ term. 
Therefore, for two target configurations $\mathcal{A}$ and $\mathcal{B}$ (e.g., different sheet sizes or materials) measured under identical conditions, i.e., at same $d$, $\lambda$, and \(\theta_b(d)\), the change in \ac{PL} due solely to the change in \ac{RCS} is given as:}
\begin{equation}\label{eq:impact_rcs_on_pl_general}
\begin{split}
\textcolor{black}{\Delta\!\operatorname{PL}^{\mathcal{B}\gets\mathcal{A}}(d,\lambda)}
&\textcolor{black}{\triangleq 
\operatorname{PL}_{\mathcal{B}}(d,\lambda) - \operatorname{PL}_{\mathcal{A}}(d,\lambda)} \\
&\textcolor{black}{= \underbrace{-10\log_{10}\!\Biggl(\frac{\sigma_{\mathcal{B}}(d,\lambda)}{\sigma_{\mathcal{A}}(d,\lambda)}\Biggr)}_{\Delta\!\overline{\operatorname{PL}}^{\mathcal{B}\gets\mathcal{A}}(d,\lambda)}
+ \Delta X_{\sigma},}
\end{split}
\end{equation}
\textcolor{black}{where $\Delta X_{\sigma} \triangleq X_{\sigma,\mathcal{B}} - X_{\sigma,\mathcal{A}}$ captures any potential difference in shadowing between the two measurement realizations.}

\textcolor{black}{In particular, in same propagation environment, the \ac{RCS} induced change reduces to the following deterministic expression:}
\begin{equation}\label{eq:impact_rcs_on_pl_deterministic}
\textcolor{black}{\Delta\!\overline{\operatorname{PL}}^{\mathcal{B}\gets\mathcal{A}}(d,\lambda)
= -10\log_{10}\!\Biggl(\frac{\sigma_{\mathcal{B}}(d,\lambda)}{\sigma_{\mathcal{A}}(d,\lambda)}\Biggr).}
\end{equation}

\textcolor{black}{Using the \ac{NF} bistatic \ac{RCS} approximations from \eqref{approx_rcs}, the impact can be written explicitly in terms of the model coefficients.
For the first-order model $\sigma_{1}(d,\lambda) = a_1\,d^{2}\cos^{m}\!\theta_b(d)$,}
\begin{equation}\label{eq:impact_sigma1}
\textcolor{black}{\Delta\!\overline{\operatorname{PL}}^{\mathcal{B}\gets\mathcal{A}}(d,\lambda)
= -10\log_{10}\!\Biggl(\frac{a_{1,\mathcal{B}}}{a_{1,\mathcal{A}}}\Biggr),}
\end{equation}
\textcolor{black}{which is a constant offset, independent of $d$ and $\lambda$, because the geometric factors $d^{2}$ and $\cos^{m}\!\theta_b(d)$ cancel in the ratio.}

\textcolor{black}{For the second-order \ac{RCS} model, i.e., $\sigma_{2}(d,\lambda)=\bigl(a_1 d^{2}+a_2\lambda d^{3}\bigr)\cos^{m}\!\theta_b(d)$, the change in the \ac{PL} can be quantified as:}
\begin{equation}\label{eq:impact_sigma2}
\textcolor{black}{\Delta\!\overline{\operatorname{PL}}^{\mathcal{B}\gets\mathcal{A}}(d,\lambda)
= -10\log_{10}\!\Bigl(
\frac{a_{1,\mathcal{B}}\,d^{2}+a_{2,\mathcal{B}}\lambda d^{3}}
{a_{1,\mathcal{A}}\,d^{2}+a_{2,\mathcal{A}}\lambda d^{3}}
\Bigr),}
\end{equation}
\textcolor{black}{and for the third-order \ac{RCS} model, $\sigma_{3}(d,\lambda)=\bigl(a_1 d^{2}+a_2\lambda d^{3}+a_3\lambda^{2}d^{4}\bigr)\cos^{m}\!\theta_b(d)$, we have:}
\begin{equation}\label{eq:impact_sigma3}
\textcolor{black}{\Delta\!\overline{\operatorname{PL}}^{\mathcal{B}\gets\mathcal{A}}(d,\lambda)
\!\!=\!\! -10\log_{10}\!\Biggl(
\frac{a_{1,\mathcal{B}}d^{2}+a_{2,\mathcal{B}}\lambda d^{3}+a_{3,\mathcal{B}}\lambda^{2}d^{4}}
{a_{1,\mathcal{A}}d^{2}+a_{2,\mathcal{A}}\lambda d^{3}+a_{3,\mathcal{A}}\lambda^{2}d^{4}}
\Biggr).}
\end{equation}

\textcolor{black}{\eqref{eq:impact_sigma2}-\eqref{eq:impact_sigma3} indicate that, unlike the constant shift of the first-order \ac{RCS} model in which the change in \ac{RCS} is due to sheet specific parameters, a change of sheet generally introduces a $d$ and $\lambda$ dependent correction to the \ac{PL} curve in addition to the sheet specific parameters. Consequently, the shape of the \(d\) trend is not preserved, because the several powers of $d$ and $\lambda$ contribute with different weights through the distinct coefficient ratios.}
\subsection{\textcolor{black}{Impact of Bistatic Angle Contribution on Path Loss}}\label{sec4d}
\textcolor{black}{In the measurement geometry considered, $\theta_b$ is not an independent variable. Fig. \ref{result_deter_rcs} highlights the dependence of \(\theta_b\) on the inverse of \ac{PL}, i.e., path gain for the considered measurement geometry. As \(\theta_b\) decreases, the path gain increases, indicating reduced propagation attenuation. The trend is consistent with the intuition that the  \ac{RCS} is typically highest for smaller bistatic separations. In addition, a quantitative relationship between \(\theta_b\) and \ac{PL} can be derived. Once the fitting parameters are obtained, the subsequent analysis can be used to evaluate the dependence of \ac{PL} on \(\theta_b\).}

\textcolor{black}{For the symmetric bistatic configuration where \(d_{\mathrm{Tx,tar}} = d_{\mathrm{Rx,tar}} = d\), $\theta_b$ at the target is a deterministic function of $d$. Accordingly, we adopt a separable \ac{RCS} model of the form:}
\begin{equation}\label{eq:sigma_factor_theta}
\textcolor{black}{\sigma_i(d,\lambda) = \bar{\sigma}_i(d,\lambda) \, \cos^{m}\!\big(\theta_b(d)\big),}
\end{equation}
\textcolor{black}{where $\bar{\sigma}_i(d,\lambda)$ contains only polynomial dependence on \(d\) and \(\lambda\), while the angular variation is captured by $\cos^{m}\theta_b(d)$. Substituting \eqref{eq:sigma_factor_theta} into the bistatic \ac{PL} expression for the \ac{FI} model in \eqref{pl_model_fi} yields:}
\begin{equation}\label{eq:pl_with_theta_of_d}
\begin{aligned}
\textcolor{black}{\mathrm{PL}_i^\mathrm{FI}(d,\lambda) }
&\textcolor{black}{= \alpha + 20n\log_{10}d 
   - 10\log_{10}\!\bigl(\bar{\sigma}_i(d,\lambda)\bigr)} \\
&\textcolor{black}{\quad - 10m\log_{10}\!\bigl(\cos\theta_b(d)\bigr) 
   + X_\sigma^\mathrm{FI}.}
\end{aligned}
\end{equation}

\textcolor{black}{For two measurement positions $d_1$ and $d_2$ with corresponding bistatic angles $\theta_b(d_1)$ and $\theta_b(d_2)$, the differential \ac{PL} for same \(\lambda\) becomes:}
\begin{equation}\label{eq:delta_pl_full_theta_of_d}
\begin{aligned}
\textcolor{black}{\Delta \mathrm{PL}_i} &\textcolor{black}{\triangleq \mathrm{PL}_i(d_2,\lambda) - \mathrm{PL}_i(d_1,\lambda)} \\
&\textcolor{black}{= 20n\log_{10}\!\left(\frac{d_2}{d_1}\right)
   - 10\log_{10}\!\left(\frac{\bar{\sigma}_i(d_2,\lambda)}{\bar{\sigma}_i(d_1,\lambda)}\right)} \\
&\textcolor{black}{\quad \underbrace{- 10m\log_{10}\!\left(\frac{\cos\theta_b(d_2)}{\cos\theta_b(d_1)}\right)}_{\Delta \mathrm{PL}_{\theta}}
   + \bigl(X_{\sigma,2} - X_{\sigma,1}\bigr).}
\end{aligned}
\end{equation}
\textcolor{black}{\eqref{eq:delta_pl_full_theta_of_d} shows explicitly that range-dependent terms do not cancel when $\theta_b$ varies with the measurement geometry. Note that, \eqref{eq:delta_pl_full_theta_of_d} is valid for both \ac{FI} and \ac{CI} (if same \(d_0\) is considered for both measurement instances) models. Moreover, for a fixed \ac{Tx}-\ac{Rx} separation $d_{\mathrm{Tx,Rx}}$, the law of cosines in the symmetric bistatic triangle gives:}
\begin{equation}\label{eq:theta_of_d_correct}
\textcolor{black}{\cos\theta_b(d) = \frac{2d^{2} - d_{\mathrm{Tx,Rx}}^{2}}{2d^{2}}
= 1 - \frac{d_{\mathrm{Tx,Rx}}^{2}}{2d^{2}},}
\end{equation}
\textcolor{black}{from which the \(\theta_b(d)\) is explicitly given as:}
\begin{equation}\label{eq:theta_of_d}
\textcolor{black}{
\theta_b(d) = \arccos\!\Bigl(1 - \frac{d_{\mathrm{Tx,Rx}}^{2}}{2d^{2}}\Bigr),}
\end{equation}

\textcolor{black}{By substituting \eqref{eq:theta_of_d} into \eqref{eq:pl_with_theta_of_d}, we can isolate the purely geometry-induced angular contribution to the \ac{PL} as:}
\begin{equation}\label{eq:pl_theta_explicit}
\textcolor{black}{\mathrm{PL}_{\theta}(d) \!\triangleq \! -10m\log_{10}\!\bigl(\cos\theta_b(d)\bigr)
\!= \!-10m\log_{10}\!\Bigl(1 - \frac{d_{\mathrm{Tx,Rx}}^{2}}{2d^{2}}\Bigr).}
\end{equation}
\textcolor{black}{The corresponding change in this angular term between two ranges \(d_1\) and \(d_2\) is given as:}
\begin{equation}\label{eq:delta_pl_theta_explicit}
\textcolor{black}{\Delta \mathrm{PL}_{\theta}
= -10m\log_{10}\!\Biggl(
\frac{ 1 - \frac{d_{\mathrm{Tx,Rx}}^{2}}{2d_2^{2}}}
     { 1 - \frac{d_{\mathrm{Tx,Rx}}^{2}}{2d_1^{2}}}
\Biggr).}
\end{equation}

\textcolor{black}{Thus, the measured \ac{PL} difference $\Delta\mathrm{PL}_i$ is fully described by the sum of three deterministic components, i.e., propagation attenuation ($20n\log_{10}(d_2/d_1)$), intrinsic \ac{RCS} range dependence, and the geometry-driven angular term $\Delta\mathrm{PL}_{\theta}$ together with the residual stochastic term $X_{\sigma,2}-X_{\sigma,1}$.}
\vspace{-2mm}
\section{Conclusions}\label{sec5}
\textcolor{black}{Our study characterizes the \ac{RCS} using statistical and deterministic modeling approaches.  In the statistical framework, we used targets including \acp{UAV}, an \ac{AMR}, and a \ac{RA} in an \ac{InF} environment to perform statistical \ac{RCS} analysis at \SI{25}{}-\SI{28}{\GHz}, where we collected \ac{RCS} measurements in both quasi-monostatic and bistatic configurations at different bistatic angles. Moreover, we compared various parametric distributions to determine the best fit for scattering properties and evaluated novel deterministic \ac{NF} \ac{RCS} models against the measured data. For the statistical \ac{RCS} characterization, we found that the target \ac{RCS} is best modeled by lognormal and gamma distributions with high confidence. Essentially, the presented results establish baseline \ac{RCS} distributions under boresight illumination for quasi-monostatic and moderate bistatic configurations. 
On the other hand, for the deterministic analysis, we used a rectangular sheet as the primary target in \ac{NF} conditions to model bistatic \ac{RCS} using \ac{PL} measurements across the \SI{25}{}-\SI{28}{\GHz} frequency range. We collected \ac{PL} measurements to derive RCS data, parameterizing the models by the bistatic angle and the relative geometry of the \ac{Tx}, \ac{Rx}, and target.
Moreover, we compared several deterministic \ac{RCS} models against the measured data, evaluating their accuracy using the \ac{MFE} and found a strong dependence of \ac{RCS} on both the bistatic angle and target distance, which helped us identify the most accurate models for characterizing \ac{NF} bistatic scattering.
For deterministic \ac{RCS} modeling, the analysis could be extended to sheets of varying sizes and materials in order to enhance the validity of the framework and validate the impact of change in \ac{RCS} on \ac{PL}. The described extensions can enhance model representativeness for edge geometries while preserving the calibrated \ac{RCS} extraction assumptions adopted in this study. Future work may also consider other frequency bands and propagation environments.}

\bibliographystyle{IEEEtran}
\bibliography{main}

\end{document}